\documentclass[a4paper,fleqn,usenatbib,amsmath,amssymb]{mnras}

\def\MyApJ#1{}
\def\MyMNRAS#1{#1}

\usepackage{graphicx}
\usepackage{bm}
\usepackage{epsf}
\usepackage{verbatim}
\usepackage{amsmath}
\usepackage{amsfonts}

\usepackage{epsfig}
\usepackage{dcolumn}
\usepackage{graphicx}
\usepackage{amssymb}
\usepackage{tablefootnote}

\usepackage{xcolor}
\definecolor{darkgreen}{rgb}{0.0,0.5,0.0}

\usepackage[percent]{overpic}

\def\myfig#1{./#1}

\def\PlotFigs#1{} 
\def\PlotFigsA#1{#1} 

\def\PaperTwo{Gurwich and Keshet, in preparation}

\MyApJ{\bibliographystyle{hapj}}
\MyMNRAS{\bibliographystyle{mnras}}

\newcommand{\ie}{\emph{i.e.,} }
\newcommand{\eg}{\emph{e.g.,} }
\newcommand{\cf}{\emph{cf.} }

\newcommand{\be}{\begin{equation}}
\newcommand{\ee}{\end{equation}}
\newcommand{\bea}{\begin{equation*}}
\newcommand{\eea}{\end{equation*}}
\newcommand{\beqr}{\begin{eqnarray} \nonumber}
\newcommand{\eeqr}{\end{eqnarray}}
\newcommand{\beqrb}{\begin{eqnarray}}
\newcommand{\eeqrb}{\nonumber \end{eqnarray}}
\newcommand{\fin}{\mbox{ .}}
\newcommand{\coma}{\mbox{ ,}}

\newcommand{\cm}{\mbox{ cm}}
\newcommand{\sr}{\mbox{ sr}}
\newcommand{\se}{\mbox{ s}}

\newcommand{\Myr}{\mbox{ Myr}}

\newcommand{\erg}{\mbox{ erg}}

\newcommand{\km}{\mbox{ km}}
\newcommand{\pc}{\mbox{ pc}}
\newcommand{\kpc}{\mbox{ kpc}}

\newcommand{\keV}{\mbox{ keV}}

\newcommand{\GeV}{\mbox{ GeV}}

\newcommand{\gama}{$\gamma$}

\newcommand{\vect}[1]{\bm{#1}}

\newcommand{\dgr}{{^\circ}}

\MyApJ{

}

\defcitealias{SuEtAl10}{S10}
\defcitealias{FermiBubbles14}{F14}

\newcommand{\Su}{{\citetalias{SuEtAl10}}}
\newcommand{\FT}{{\citetalias{FermiBubbles14}}}

\defcitealias{KeshetGurwich16_Diffusion}{KG16}

\MyApJ{
\begin{document}}

\MyApJ{
\title{Fermi bubbles: high latitude X-ray supersonic shell}
\shorttitle{X-ray Fermi-bubble shell}
}

\MyApJ{
\author{Uri Keshet\altaffilmark{1} and Ilya Gurwich\altaffilmark{2}}
\shortauthors{Keshet \& Gurwich}
}



\MyMNRAS{
\title[X-ray Fermi-bubble shell]{Fermi bubbles: high latitude X-ray supersonic shell}
\author[Keshet \& Gurwich]{
Uri Keshet$^{1}$\thanks{E-mail: ukeshet@bgu.ac.il}
and Ilya Gurwich$^{2}$
\\
$^{1}$Physics Department, Ben-Gurion University of the Negev, POB 653, Be'er Sheva 84105, Israel\\
$^{2}$Department of Physics, NRCN, POB 9001, Be'er Sheva 84190, Israel
}
}

\MyMNRAS{
\pubyear{2017}
\begin{document}
\label{firstpage}
\pagerange{\pageref{firstpage}--\pageref{lastpage}}
\maketitle
}

\MyApJ{\date{\today}}

\begin{abstract}
The nature of the bipolar, $\gamma$-ray Fermi bubbles (FB) is still unclear, in part because their faint, high-latitude X-ray counterpart has until now eluded a clear detection.
We stack \emph{ROSAT} data at varying distances from the FB edges, thus boosting the signal and identifying an expanding shell behind the southwest, southeast, and northwest edges, albeit not in the dusty northeast sector near Loop I.
A Primakoff-like model for the underlying flow is invoked to show that the signals are consistent with halo gas heated by a strong, forward shock to $\sim$keV temperatures.
Assuming ion--electron thermal equilibrium then implies a $\sim10^{56}$ erg event near the Galactic centre $\sim7$ Myr ago.
However, the reported high absorption-line velocities suggest a preferential shock-heating of ions, and thus more energetic ($\sim 10^{57}$ erg), younger ($\lesssim 3$ Myr) FBs.
\end{abstract}

\MyMNRAS{
}

\MyApJ{}

\MyApJ{\maketitle}

\section{Introduction}
\label{sec:Intro}

\subsection{The Fermi bubbles}

Non-thermal lobes emanate from the nuclei of many galaxies.
These lobes, thought to arise from starburst activity or an outflow from a super-massive black hole \citep[for reviews, see][]{VeilleuxEtAl05,KingPounds15}, play an important role in the theory of galaxy formation \citep[\eg][and references therein]{Benson10}.
The presence of a massive, bipolar outflow in our own Galaxy has long been suspected, largely based on X-ray and radio signatures on large \citep{Sofue00}, intermediate \citep{BlandHawthornAndCohen03} and small \citep{BaganoffEtAl03} scales, indicating an energetic, $\gtrsim 10^{55}\erg$ event \citep[][and references therein]{VeilleuxEtAl05}.

This scenario was revived by the discovery \citep[][henceforth \Su]{DoblerEtAl10, SuEtAl10} of two large \gama-ray, so called Fermi, bubbles (FBs), symmetrically rising far from the Milky-Way plane yet morphologically connected, at least approximately and in projection, to the intermediate scale X-ray outflow features.
Due to their dynamical, nonthermal nature, and the vast energy implied by their presumed Galactic-scale distance, an accurate interpretation of the FBs is important for understanding the energy budget, structure, and history of our Galaxy.

The FBs, extending out to latitudes  $|b|\simeq 55^\circ$,
are also seen in microwave synchrotron emission \citep{Dobler12, PlanckHaze13}, as the so-called microwave Haze \citep{Finkbeiner04}, a residual diffuse signal surrounding the Galactic centre (GC).
They may also morphologically coincide and with linearly polarized radio emission \citep{CarrettiEtAl13}, although the association of this signal with the FBs is unclear.

\subsection{Interpretation as a Galactic-scale phenomenon}

The tentative identification of the FBs as massive structures emanating from the GC, rather than small lobes of a nearby object seen in projection, was based mainly on the coincidence of the lobe's base, to within a few degrees, with the GC.
The FB edges were recently extracted robustly, without making any assumptions concerning the Galactic foreground, by applying gradient filters to the Fermi-LAT map; the resulting edges connect smoothly to the intermediate-low latitude X-ray features, strengthening the FB--GC coincidence to sub-degree scales \citep[][henceforth \citetalias{KeshetGurwich16_Diffusion}]{KeshetGurwich16_Diffusion}.
The implied, $\sim 10\kpc$ distance scale corresponds to a high, $\sim 4\times 10^{37}\erg\se^{-1}$ luminosity (\Su, \FT).

Additional, less direct indications for the FB--GC association include the radio emission being too faint for a nearby source confined to the already-magnetized thick Galactic disk, the orientation of the FB axis being nearly exactly perpendicular to the Galactic plane, in agreement with an extended structure bursting out of the Galactic disk \citepalias[][and references therein]{KeshetGurwich16_Diffusion}, and the low-latitude depolarization of the tentatively associated linearly polarized lobes \citep{CarrettiEtAl13}.
Another claim is the fairly high emission measure (EM) of possibly related high-latitude X-ray features \citep{KataokaEtAl13}; however, we argued in \citetalias{KeshetGurwich16_Diffusion} and conclusively show here that the X-ray signature was incorrectly interpreted.
Moreover, it is unclear if $\mbox{EM}\simeq 0.01\cm^{-6}\pc$ suffices to rule out a local structure.

\subsection{Underlying flow and edge shock}

In spite of their dramatic appearance in the \gama-ray sky, the nature of the FBs is still debated.
Different models were proposed (\Su, \FT), interpreting
the FB edge as an outgoing shock \citep{FujitaEtAl13}, a termination shock of a wind \citep{Lacki14, MouEtal14}, or a discontinuity \citep{Crocker12, GuoMathews12, SarkarEtAl15};
the \gama-ray emission mechanism as either hadronic \citep{CrockerAharonian11, FujitaEtAl13} or leptonic \citep{YangEtAl13};
the underlying engine as a starburst \citep{CarrettiEtAl13, Lacki14, SarkarEtAl15}, a jet from the the central massive black hole \citep{ChengEtAl11, GuoMathews12, ZubovasNayakshin12, MouEtal14}, or steady star-formation \citep{Crocker12};
and the cosmic-ray (CR) acceleration mechanism as first order Fermi acceleration, second order Fermi acceleration \citep{MertschSarkar11, ChernyshovEtAl14}, or injection at the GC \citep{GuoMathews12, Thoudam13}.

More clues regarding the nature of the FBs have gradually surfaced.
The microwave haze shows a hard spectrum, $\nu I_\nu\propto \nu^{-0.55\pm0.05}$ \citep[\eg][]{PlanckHaze13}, corresponding to CR electron (CRE) acceleration in a strong, $M\gtrsim5$ shock \citepalias{KeshetGurwich16_Diffusion}.
Metal absorption lines of $\{-235,+250\}\km \se^{-1}$ line of sight velocities in the spectrum of quasar PDS 456, located near the base of the northern FB, indicate an outflow velocity $\gtrsim 900\km \se^{-1}$ \citep{FoxEtAl15}.
Longitudinal variations in the
\MyMNRAS{\ion{O}{VII}}
\MyApJ{\ion{O}{7}}
and
\MyMNRAS{\ion{O}{VIII}}
\MyApJ{\ion{O}{8}}
emission line strengths, integrated over a wide latitude range covering the entire FBs, suggest a $\sim 0.4\keV$ FB multiphase plasma with a denser, slightly hotter edge, propagating through a $\sim 0.2\keV$ halo, thus suggesting a forward shock of Mach number $M=2.3_{-0.4}^{+1.1}$ \citep{MillerBregman16}.
By removing a FB template, \citet{SuFinkbeiner12} found a southeast--northwest, bipolar jet, with a cocoon on its southeast side; however, only the cocoon was so far confirmed to be significant (\FT).

The \gama-ray spectrum of the FB shows very little variations with position along the edge; this alone indicates, when invoking CRE Fermi-acceleration, a strong shock with $M>5$ \citepalias{KeshetGurwich16_Diffusion}.
The spatially integrated \citepalias{SuEtAl10,FermiBubbles14} or locally measured \citepalias{KeshetGurwich16_Diffusion} \gama-ray spectrum can be naturally explained for $\sim$few Myr old bubbles, without invoking ad-hoc energy cutoffs, only in a leptonic model featuring a $\sim 1\GeV$ cooling break (\PaperTwo); this spectrum is again consistent with CREs injected in a strong shock.
Finally, the edge spectrum is found to be slightly but uniformly and consistently softer than the FB-integrated spectrum \citepalias{KeshetGurwich16_Diffusion}; this is naturally explained by inward CRE diffusion in a Kraichnan-like magnetic turbulence if the FB edge is a forward shock.

\subsection{High latitude X-ray signature}

At the highest FB latitudes, an absorbed, $\sim0.3\keV$ X-ray component was reported \citep{KataokaEtAl13}, with a $\sim 60\%$ jump in the EM as one crosses outside the edge of the northern bubble.
Here and in what follows, we refer to the gas closer to (farther from) the GC as lying below, or equivalently inside (above, or equivalently outside) the edge.
The putative jump reported by \citet{KataokaEtAl13} would suggest that the FB edges are in fact a weak, $M\sim1.5$ reverse shock, terminating a wind.
However, as we pointed out in \citetalias{KeshetGurwich16_Diffusion}, these observations are complicated by the high level of dust and confusion with other structure in the northern hemisphere, and are equally --- if not more convincingly --- consistent with a drop, rather than a jump, in both south and north bubbles, which would suggest a forward shock.
Such a drop would furthermore be consistent with the evident X-ray drops at the intermediate-low latitude X-ray features \citep{BlandHawthornAndCohen03}, and with the other evidence outlined above.

Here we use the \emph{ROSAT} all sky survey \citep[RASS;][]{SnowdenEtAl97} to measure the high latitude X-ray signal associated with the FBs.
Modelling the FBs as an expanding shell, we derive the drop in flux and in temperature expected as one crosses outside the edge, in a transition spanning $\sim 2\dgr$ in projection, as well as the gradual brightening of the signal and cooling of the gas over $\sim10\dgr$ within the FBs towards the GC.
These signals are difficult to pick up directly due to various structure in the X-ray sky, the uncertainty in the precise location of the FB edge, the low surface brightness, and the \emph{ROSAT} X-ray background.
Nevertheless, the data can be stacked at fixed distances from the FB edge, which has already been traced directly with a few degree precision using the Fermi data in \citetalias{KeshetGurwich16_Diffusion}, to greatly enhance the signal well beyond the detection threshold.
Errors in edge location, variations in the radial profiles, and noise, are thus effectively averaged out, enabling a firm detection of the signal.

The paper is arranged as follows.
In \S\ref{sec:Model} we model the expected X-ray signal from the FBs.
The \emph{ROSAT} data and analysis procedure are presented in \S\ref{sec:Data}.
The X-ray structure perpendicular to the FB edges is extracted in \S\ref{sec:RadialProfiles}, and analyzed in \S\ref{sec:Analysis}.
The results are summarised and discussed in \S\ref{sec:Discussion}.
We use $1\sigma$ error bars, unless otherwise stated, and Galactic coordinates, throughout.

\section{FB model}
\label{sec:Model}

\subsection{Edge toy model}

To model the gas flow underlying the FBs, we begin with a toy model for the shape of the high latitude edge.
Consider a simple bipolar shock pattern, specified by the Galactocentric radius,
\begin{equation} \label{eq:FBEdgeFull}
R = R_0 \times \begin{cases}
1-\left(\theta/\theta_0\right)^2 & \mbox{for $0<\theta<\theta_1$ ;}\\
c_1/(\theta+\theta^8) & \mbox{for $\theta_1<\theta<\pi/2$}
\end{cases}
\end{equation}
for the north FB, and symmetrically $(\theta\to\pi-\theta$) about the Galactic plane for the south FB.
Here, $R_0$ is the peak height of the FB, and $\theta$ is the polar angle, measured in a frame with the GC at the origin.
For a FB seen in projection with a maximal latitude $b\simeq 53\dgr$, and for a Solar Galactocentric radius $r_\odot\simeq8.5\kpc$, we find that $R_0\simeq 10\kpc$.
A good fit for the top of the FBs requires $\theta_0\simeq \pi/5$; continuity of the edge and its first derivative then yield $\theta_1\simeq\pi/8.6$ and $c_1\simeq 0.24$.
The FB edge resulting from this two-parameter ($R_0$ and $\theta_0$) model is shown in projection in Figure \ref{fig:FBModel}.

For comparison, the figure also shows the FB edges, as extracted in \citetalias{KeshetGurwich16_Diffusion} (coarse grained edge number 1, therein and below), for both hemispheres.
At latitudes $|b|\gtrsim 15\dgr$, the model reasonably matches the observed edge on the better resolved, eastern side; lower latitudes are outside the scope of the present analysis.
The model does not, however, capture the east-west FB asymmetry, in particular the observed westward extension of the high $|b|$ bubbles.
Consequently, the high latitude ($|b>15\dgr$) solid angle $\sim 0.34$ of each observed FB is $\sim20\%$ larger than the corresponding, $\sim 0.28\sr$ solid angle of the FB in the toy model. 

\subsection{Upstream, halo model}

Consider a scenario where the FBs arise from a rapid release of energy, leading to a supersonic outflow with a forward shock coincident with the FB edges.
The outflowing gas should form a massive shell, with density and pressure increasing outwards towards the shock.
This increase is expected to be gradual for the $\sim r^{-2}$ decline attributed to the density of the Galaxy's hot gas halo, into which the FBs are presumably expanding.

In particular, a $\beta$-model based on
\MyMNRAS{\ion{O}{VII}}
\MyApJ{\ion{O}{7}}
emission and absorption lines \citep{MillerBregman13} is consistent at $r\gg1\kpc$ radii with an isothermal sphere distribution,
\begin{equation}\label{eq:nu}
n_{e,u} \simeq  n_{e,10} \left( \frac{r}{10\kpc} \right)^{-2} \coma
\end{equation}
where $n_{e,10}\equiv 4\times 10^{-4}n_{4}\cm^{-3}$ is the electron number density $n_e$ at $r=10\kpc$, and subscript $u$ (subscript $d$) denotes the upstream (downstream) plasma.

Specifying the flow underlying the FBs requires some assumption on the upstream temperature, $T_u$.
The halo temperature, based on
\MyMNRAS{\ion{O}{VII}}
\MyApJ{\ion{O}{7}}
emission and absorption, is \citep{MillerBregman13}
\begin{equation} \label{eq:Tu}
k_B T_u \equiv k_B T_h\simeq (0.1\mbox{--}0.2) \keV  \coma
\end{equation}
where $k_B$ is Boltzmann's constant.
Assuming (henceforth) an adiabatic index $\Gamma=(5/3)$ and a cosmic element abundance with mean particle mass $\mu m_p$, where $\mu\simeq 0.6$, these temperatures correspond to an upstream sound velocity $c_{s,u}\simeq (170\mbox{--}190) \km \se^{-1}$.
Note that a somewhat higher, $k_B T_u\simeq 0.3\keV$ temperature was derived from X-rays using \emph{Suzaku} \citep{KataokaEtAl13}.

\MyApJ{\begin{figure}[h]}
\MyMNRAS{\begin{figure}}
\PlotFigsA{
\epsfxsize=8.5cm \epsfbox{\myfig{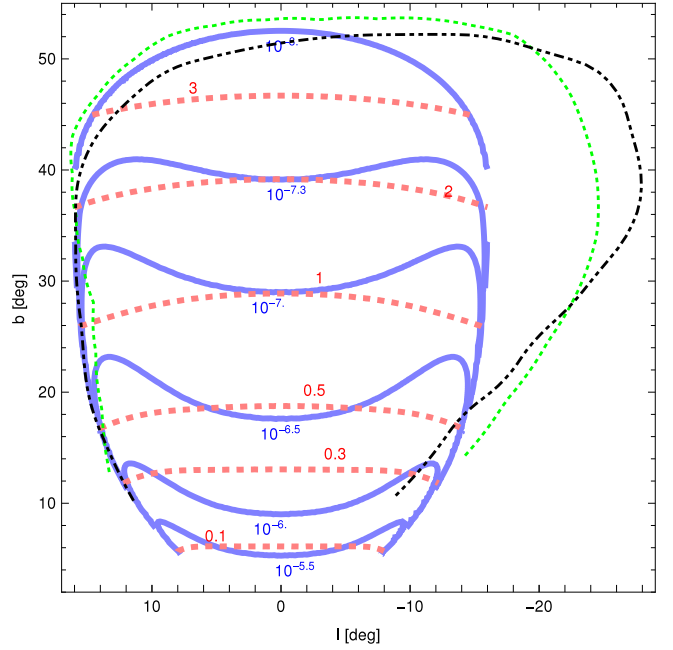}}
}
\caption{
Projected X-ray FB model: flux in the \emph{ROSAT} (0.1--2.4 keV) band (solid blue curves), $\log_{10}(F_X [n_{4}^2\erg\se^{-1}\cm^{-2}\sr^{-1}])$, and X-ray weighted temperature (dashed red contours), $T_X[M_{10}^2T_{0.15}\keV]$.
Also shown are edges 1 of the north (dot-dashed black) and (reflected about the Galactic plane) south (dotted green) FBs.
\label{fig:FBModel}
}
\end{figure}

\subsection{Flow model}

A spherical, strong shock propagating into a halo-like, $n\propto r^{-2}$ medium asymptotes to the Primakoff-like solution \citep{CourantFriedrichs48, Keller56}, in which the downstream distribution follows power-law profiles \citep{BernsteinBook80}
\begin{equation} \label{eq:PrimakoffScaling}
n\propto r \mbox{ ,}\quad P\propto r^3 \mbox{ ,}\quad\mbox{and}\quad v\propto r \fin
\end{equation}

For simplicity, we assume that the linear (in $r$) velocity of the spherical Primakoff-like solution remains valid in our nonspherical model, when choosing the GC as the origin. Then \citep{BernsteinBook80}
\begin{equation} \label{eq:LinearVScaling}
\vect{v}(\vect{r},t)= (2t)^{-1}\vect{r} \coma
\end{equation}
where $t$ is the age of the FBs.
For simplicity, we assume here that electrons and ions are shock-heated to the same temperature, and revisit this issue in \S\ref{sec:Discussion}.
These assumptions now fix the structure of the flow throughout the modelled FBs.

The Rankine-Hugoniot jump conditions dictate that the shock velocity in the Galaxy frame is
\begin{equation}
v_s=v_u=\frac{\Gamma+1}{\Gamma-1+2M^{-2}} v_d \simeq \frac{4}{1+3M^{-2}} v_d \coma
\end{equation}
where $v_{u}$ and $v_{d}$ are the fluid velocities with respect to the shock.
Taking the velocities as radial, $v_d\simeq v_s-v(R)$, so
\begin{equation}
v_s = M c_{s,u} \simeq \frac{4v(R)}{3(1-M^{-2})} = \frac{2R/t}{3(1-M^{-2})}  \coma
\end{equation}
giving
\begin{equation}
M=\frac{2}{3}\tilde{M}+\sqrt{1+\left(\frac{2}{3}\tilde{M}\right)^2} \coma
\end{equation}
where $\tilde{M}\equiv v(R)/c_{s,u}=R/(2c_{s,u}t)$ is the fluid velocity at $r=R$ normalized to upstream sound.
In the strong shock limit, this becomes $M\simeq(4/3)\tilde{M}$.

Equivalently, given a shock Mach number $M$, the flow velocity is fixed by $v(R)=3c_{s,u}(M-M^{-1})/4$, becoming $v(R)\simeq (3/4)Mc_{s,u}$ in the strong shock limit.
We may now write $t$ in terms of the Mach number $M=10 M_{10}$ at the top of the FB,
\begin{equation}
t = \frac{2R_0/(3c_{s,u})}{M-M^{-1}} \simeq \frac{2R_0}{3Mc_{s,u}} \simeq 3.3 M_{10}^{-1}T_{0.15}^{-1/2} \Myr \coma
\end{equation}
where $T_{0.15}\equiv k_B T_u/(0.15\keV)$, and in the second equality we assumed the shock to be strong.

The thermal properties in the immediate downstream are now given by
\begin{equation} \label{eq:NScaling}
n_{e,d}=\frac{\Gamma+1}{\Gamma-1+2M^{-2}} n_{e,u} \simeq \frac{4}{1+3M^{-2}} n_{e,u} \coma
\end{equation}
which scales as $n_{e,u}\propto R^{-2}$ in the strong shock limit, and
\begin{equation}\label{eq:PScaling}
P_{e,d}=\left(\frac{2\Gamma M^2}{\Gamma+1}-\frac{\Gamma-1}{\Gamma+1}\right) P_{e,u} \simeq \frac{5M^2-1}{4} P_{e,u} \coma
\end{equation}
which for a strong shock becomes $\propto v(R)^2 P_{e,u}\propto R^0$, interestingly implying a constant downstream pressure along the entire shock surface at any given time.

\subsection{X-ray signature}

By combining the edge pattern in Eq.~(\ref{eq:FBEdgeFull}), the flow profiles in Eqs.~(\ref{eq:PrimakoffScaling}--\ref{eq:LinearVScaling}), the jump conditions in Eqs.~(\ref{eq:NScaling}--\ref{eq:PScaling}), and the upstream distribution in Eqs.~(\ref{eq:nu}--\ref{eq:Tu}), we may now compute the expected X-ray signature of the FBs.
The properties of the resulting signal, as seen in projection from the Solar system, are shown in Figures \ref{fig:FBModel} and \ref{fig:FBModelProj}.
They depend on the shock and upstream parameters, calibrated for simplicity at the top of the FB; in particular, we use the Mach number $M(r=10\kpc)=10M_{10}$ and electron number density $n_e(r=10\kpc)=4\times 10^{-4} n_4 \cm^{-3}$.

For comparison with the \emph{ROSAT} data analyzed in \S\ref{sec:Data}--\ref{sec:Analysis}, we model the full \emph{ROSAT}, $(0.1\mbox{--}2.4)\keV$ band.
The emission coefficient integrated over this enegry range, computed using the MEKAL model
\citep{MeweEtAl85, MeweEtAl86, Kaastra92, LiedahlEtAl95} in XSPEC v.12.5 \citep{Arnaud96}, does not strongly depend on $\sim \keV$ temperature,
\begin{equation} \label{eq:FxMEKAL}
j_{X}\simeq 9\times 10^{-25}\frac{n_{e}^2 Z_{0.3}^{0.6}}{T_{keV}^{0.1}}\erg\se^{-1}\cm^{-3}\sr^{-1} \coma
\end{equation}
where $T_{keV}\equiv(k_B T_e/1\keV)$ is the electron temperature, $Z_{0.3}\equiv Z/(0.3Z_\odot)$ is the metallicity, and the fit pertains to the $T_{keV}\in[0.1,1.5]$, $Z_{0.3}\in[0.3,3]$ range.

Neglecting for simplicity the temperature and metallicity dependencies, we integrate the approximate $j_X\simeq C_X n_e^2$ along the line of sight $l$,
\begin{align}
F_{X} & \simeq \int C_X n^2 \,dl \simeq C_X \int \left(\frac{r}{R}n_{e,d}\right)^2 \, dl \nonumber \\
& \simeq \frac{16C_X n_{e,10}^2}{(1+3M^{-2})^2} \int \frac{(10\kpc)^4 r^2}{R^6} \, dl \fin
\end{align}
Here, $R=R(\vect{r})$ is the shock radius along the ray emanating from the GC and passing through $\vect{r}$.
The X-ray-weighted temperature can similarly be computed,
\begin{align}
T_X & \simeq 
\frac{C_X}{F_X} \int \left(\frac{r^2}{R^2}T_{e,d}\right) \left(\frac{r}{R}n_{e,d}\right)^2 \, dl \nonumber \\
& \simeq \frac{500 M_{10}^2 C_X n_{e,10}^2 T_h}{F_X} \int \frac{(10\kpc)^2 r^4}{R^6} \, dl  \nonumber \\
& \simeq \frac{31M_{10}^2 T_h}{(10\kpc)^2} \, \frac{\int (r^4/R^6) \, dl}{\int (r^2/R^6) \, dl} \coma
\end{align}
where in the last two lines we approximated the shock as strong.
The resulting, projected X-ray structure, as seen by a putative observer in the Solar system, assumed to be $r_\odot=8.5\kpc$ from the GC, is shown in Figure \ref{fig:FBModel}.

In \S\ref{sec:RadialProfiles} we measure the \emph{ROSAT} flux as a function of the angular distance $\psi$ from the FB edge.
For consistency with \citetalias{KeshetGurwich16_Diffusion}, we take $\psi<0$ to designate regions inside the bubble.
To boost the signal and allow such a measurement, we stack data along wide sectors, in particular sectors defined by intermediate ($15\dgr<|b|<30\dgr$) or high ($|b|>30\dgr$) latitudes.
In order to compare these results with the model, we apply the same procedure to the modelled X-ray signature in Figure \ref{fig:FBModel}; the resulting profiles are shown in Figure \ref{fig:FBModelProj}.
Figures  \ref{fig:FBModel} and  \ref{fig:FBModelProj} show both $F_X$ (solid) contours and $T_X$ (dashed) contours.

\MyApJ{\begin{figure}[h]}
\MyMNRAS{\begin{figure}}
\PlotFigsA{
\epsfxsize=8.5cm \epsfbox{\myfig{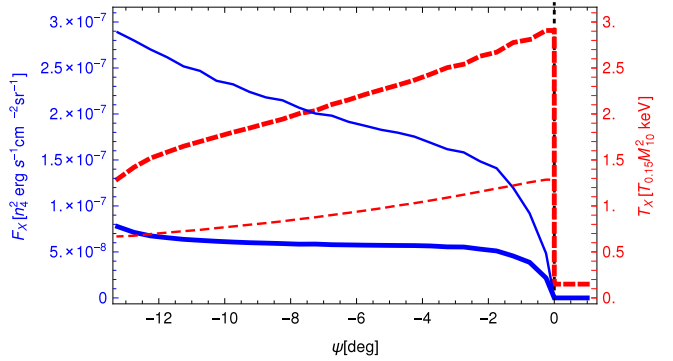}}
}
\caption{
Stacked X-ray signature of the FB model as a function of angular distance $\psi$ from the edge (marked by a dotted black line): flux $F_X$ in the \emph{ROSAT} band (solid blue curves, left axis) and X-ray weighted temperature (dashed red, right axis), shown for both the high latitude ($|b|>30\dgr$, thick curves) and intermediate latitude ($15\dgr<|b|<30\dgr$, thin curves) sectors.
\label{fig:FBModelProj}
}
\end{figure}

\subsection{Other model properties}

The total energy in the modelled FBs is estimated by integrating the ion bulk kinetic energy and the thermal energy, which we temporarily assume to be equilibrated between ions and electrons (see \S\ref{sec:Discussion} for a discussion of this assumption).
In the above model, this yields
\begin{equation}
E_{FB} \simeq 2.4 \times 10^{56} M_{10}^2n_4 T_{0.15} \erg
\end{equation}
from the two bubbles combined; $\sim 45\%$ of this energy is in the form of bulk kinetic energy.

We confirm that in our Primakoff-like model, the \gama-ray signature is broadly consistent with Fermi observations.
It rises across several degrees from the edge inward, remains quite spatially flat, and shows no limb brightening, in agreement with the data.
This is demonstrated in Figure \ref{fig:FBModelProjGamma}, showing the profiles of the \gama-ray flux in the high and intermediate latitude sectors.
The figure depicts both $j_\gamma \propto n_e^0$ and $j_\gamma \propto n_e^1$ \gama-ray emissivity models (each with its own arbitrary units); the former corresponds to the strong diffusion limit.
Interestingly, the $\propto n_e^1$ model provides a better fit to the results of \citetalias{KeshetGurwich16_Diffusion}, in which the stacked, low-energy signal appears to be stronger at lower latitudes than it is in high latitudes.

\MyApJ{\begin{figure}[h]}
\MyMNRAS{\begin{figure}}
\PlotFigsA{
\epsfxsize=8.5cm \epsfbox{\myfig{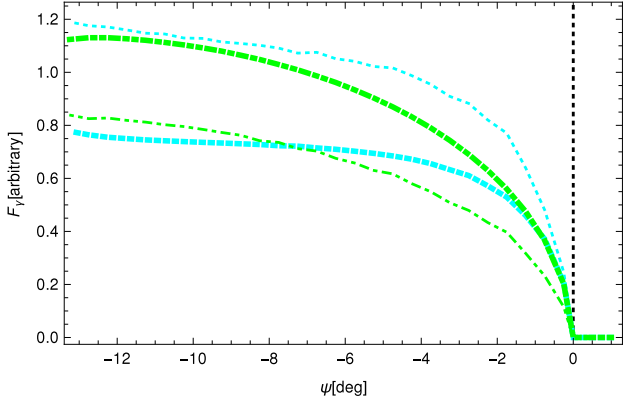}}
}
\caption{
Stacked $\gamma$-ray signature of the FB model for emissivities $j_\gamma\propto n_e^0$ (green, dot-dashed curves) and $j_\gamma\propto n_e^1$ (cyan, dotted), in the high (thick) and intermediate (thin) latitude sectors. An arbitrary normalization is applied for each emissivity model.
\label{fig:FBModelProjGamma}
}
\end{figure}

In our model, quasar PDS 456 would show $\sim\{-50,+200\}M_{10}T_{0.15}^{1/2}\km\se^{-1}$ absorption line velocities in the Galactic standard of rest (GSR), due to the FBs.
For our fiducial parameters, these offsets are somewhat smaller than, namely are only $\sim\{1/4,2/3\}M_{10}T_{0.15}^{1/2}$ times, the $\{-190,+295\}\km\se^{-1}$ GSR line velocities inferred from observations \citep{FoxEtAl15}.
The observed line velocities, and importantly, their ratio, are not well-reproduced here, but these values are sensitive to the assumed linear velocity and the adopted edge pattern at low latitudes, which are not well constrained.
Indeed, a $2.5\dgr$ eastward shift in the position of the $b=10\dgr$ modelled east FB edge would reproduce the observed values for $M_{10}^2T_{0.15}=1$.

Our analysis can be readily generalized for different choices of the 3D FB edge, flow, and upstream models.
As an example, consider the case where the upstream density is constant, giving rise to a Sedov-Taylor-Von Neumann-like profile behind the shock.
Here, the mass shell is more compact, compressed against the shock.
The resulting X-ray and \gama-ray profiles are shown in Figure \ref{fig:ST}, for a fixed upstream density $n_e=4\times 10^{-4}n_4\cm^{-3}$.
The projected profiles of X-rays and of $j_\gamma\propto n_e^1$ \gama-rays show clear limb brightening.
As we show, such profiles are inconsistent with both \emph{ROSAT} and Fermi-LAT observations; the X-ray data thus favor the standard, $n_e\propto r^{-2}$ upstream profile of the Primakoff-like model.

\MyApJ{\begin{figure}[h]}
\MyMNRAS{\begin{figure}}
\PlotFigsA{
\epsfxsize=8.5cm \epsfbox{\myfig{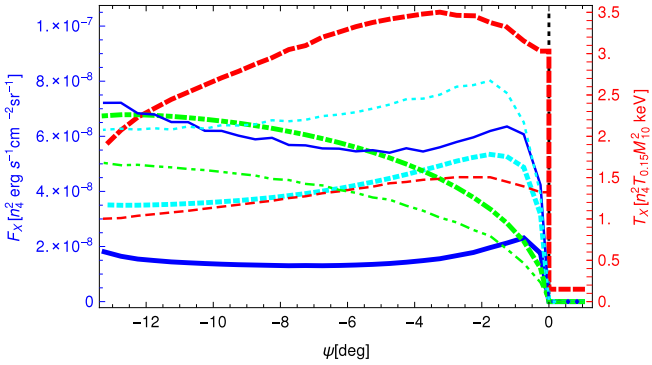}}
}
\caption{
Stacked X-ray and $\gamma$-ray signatures of the FB for an underlying Sedov-Taylor-Von Neumann-like flow.
Notations and symbols combine those of Figures \ref{fig:FBModelProj} and \ref{fig:FBModelProjGamma}.
\label{fig:ST}
}
\end{figure}

\section{Data preparation and analysis}
\label{sec:Data}

We use the \emph{ROSAT} all sky survey \citep[RASS;][]{SnowdenEtAl97}, with the Position Sensitive Proportional Counter (PSPC) of the X-ray telescope (XRT).
The provided\footnote{http://hea-www.harvard.edu/rosat/rsdc.html} PSPC maps were binned onto $12'\times12'$ pixels, well above the native $1'.8$ radius for $50\%$ energy containment.
Point sources were removed to a uniform source flux threshold for which their catalog is complete over $90\%$ of the sky; the corresponding pixels are masked from our analysis.
We use the four high energy bands of the survey, denoted R4--R7, spanning the energy range $0.4\mbox{--}2.0\keV$ with a considerable overlap, as detailed in Table \ref{tab:ROSATBands}.
The low energy bands, R1--R3, are found to be too noisy for our analysis.

Figure \ref{fig:EdgesROSAT6} illustrates the analysis using the R6 band.
The figure, spanning $160\dgr$ in latitude and $70\dgr$ in longitude, was retrieved from SkyView \citep{McGlynnEtAl98} in a rectangular (CAR) projection with a $16'$ latitude resolution, chosen to slightly exceed the native, binned map resolution.
A biconal, heart shaped signature reminiscent of the model in Figure \ref{fig:FBModel} is evident at the base of the FBs, at low latitudes $|b|\lesssim 15\dgr$ \citep{BlandHawthornAndCohen03}.

However, this emission is seen to extend to higher latitude, at least in the southern bubble, as we demonstrate by smoothing the map on large scales (for illustrative purposes only; no smoothing is used in the subsequent analysis).
For example, using an $8\dgr$ Gaussian filter (panels $c$ and $d$) shows that the signal (highlighted as long-dashed yellow contours in panel $d$) extends to $|b|\simeq 40\dgr$ latitudes.
The signal is less clear in the northern bubble, which is known to be more contaminated \citepalias[\eg][]{FermiBubbles14} due to higher levels of dust and gas \citep[\eg][]{NarayananSlatyer16}, especially near the northeastern Loop I feature; the signal may nevertheless be discernible in its northwestern part.

\MyApJ{\begin{table}[h]}
\MyMNRAS{\begin{table}}
\begin{center}
\caption{\label{tab:ROSATBands} \emph{ROSAT} energy bands used in the analysis.
}
\begin{tabular}{cc}
\hline
Band name & Energy range [keV; $10\%$ of peak response] \\
\hline
R4 & 0.44--1.01 \\
R5 & 0.56--1.21 \\
R6 & 0.73--1.56 \\
R7 & 1.05--2.04 \\
\hline
\end{tabular}
\MyApJ{\tablenotetext{1}{$10\%$ of peak response; keV units.}}
\end{center}
\end{table}

\MyApJ{\begin{figure*}[t]}
\MyMNRAS{\begin{figure*}}
\PlotFigsA{
\centerline{
\begin{overpic}[width=4.86cm]{\myfig{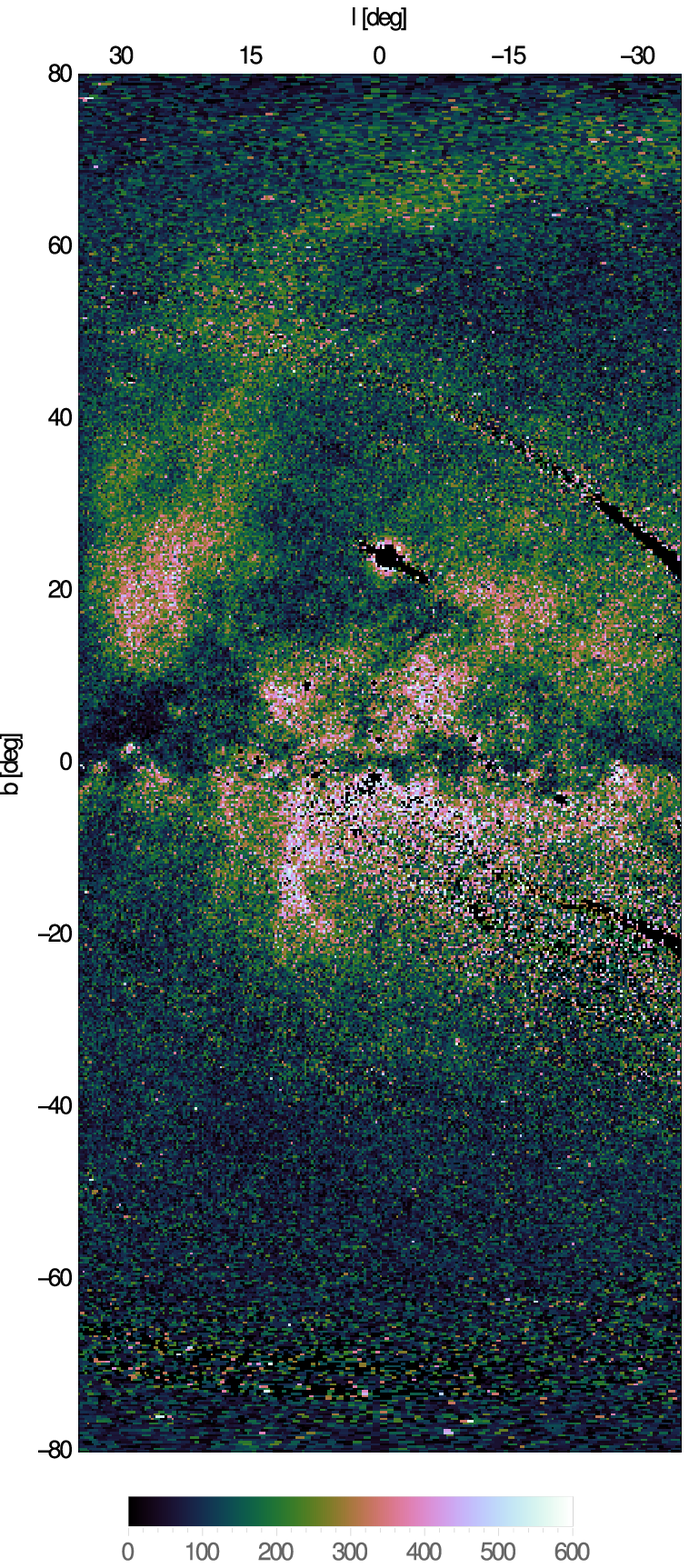}}
 \put (6,91) {\normalsize \textcolor{white}{(a)}}
\end{overpic}
\begin{overpic}[width=4.3cm, trim=0 -0.29cm 0 0]{\myfig{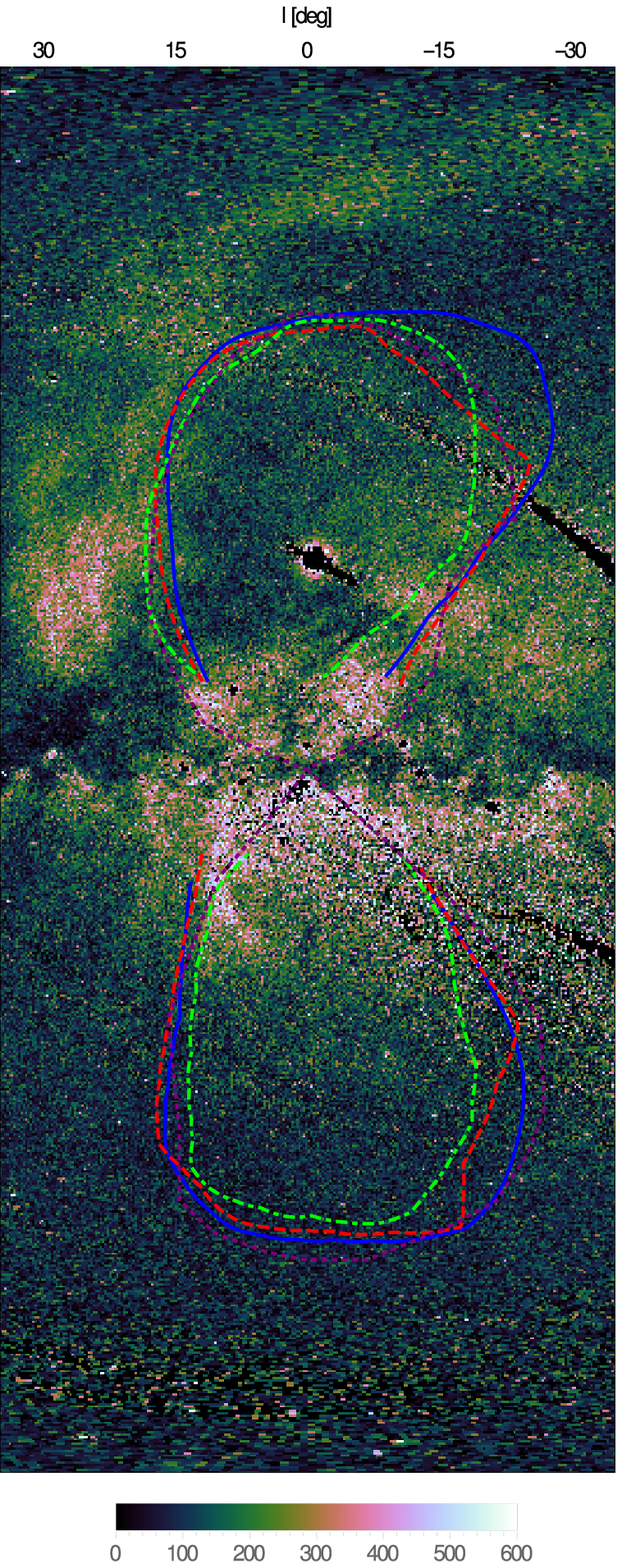}}
 \put (1,92) {\normalsize \textcolor{white}{(b)}}
\end{overpic}
\begin{overpic}[width=4.3cm, trim=0 -0.29cm 0 0]{\myfig{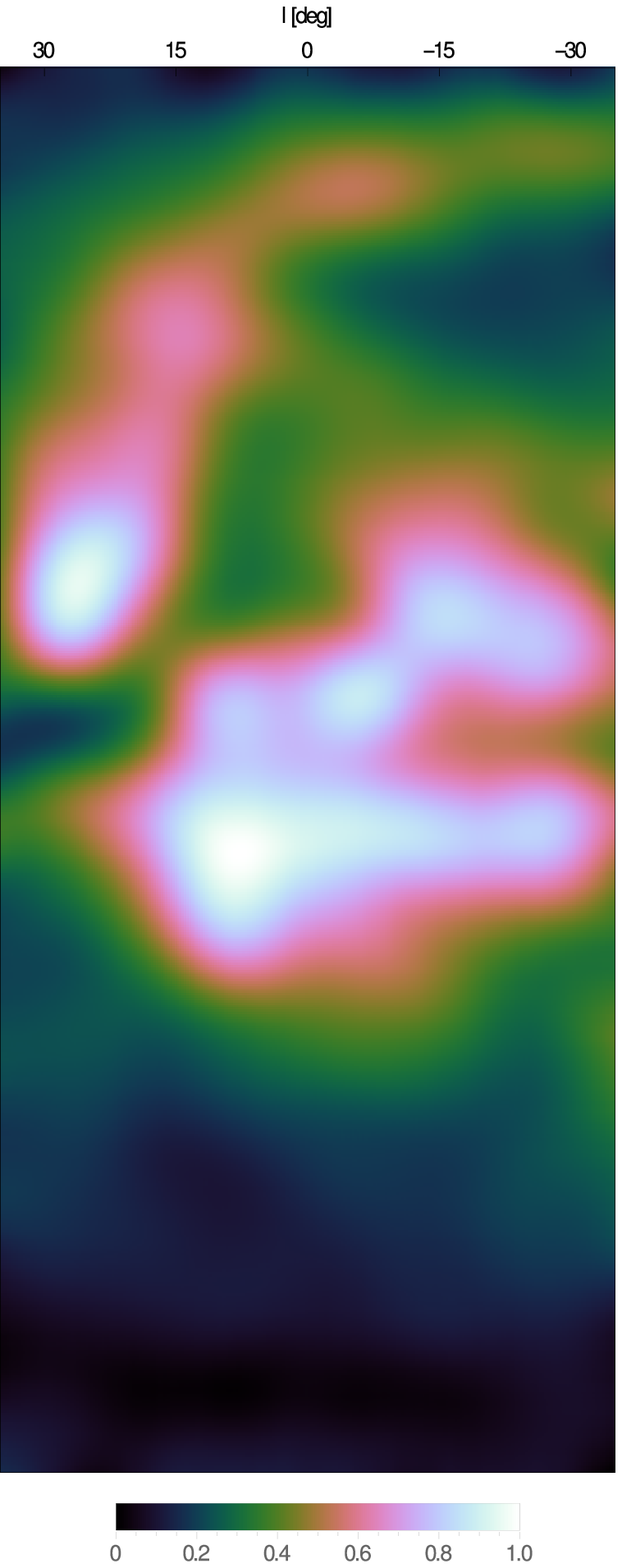}}
 \put (1,92) {\normalsize \textcolor{white}{(c)}}
\end{overpic}
\begin{overpic}[width=5.05cm, trim=0 0.15cm 0 0]{\myfig{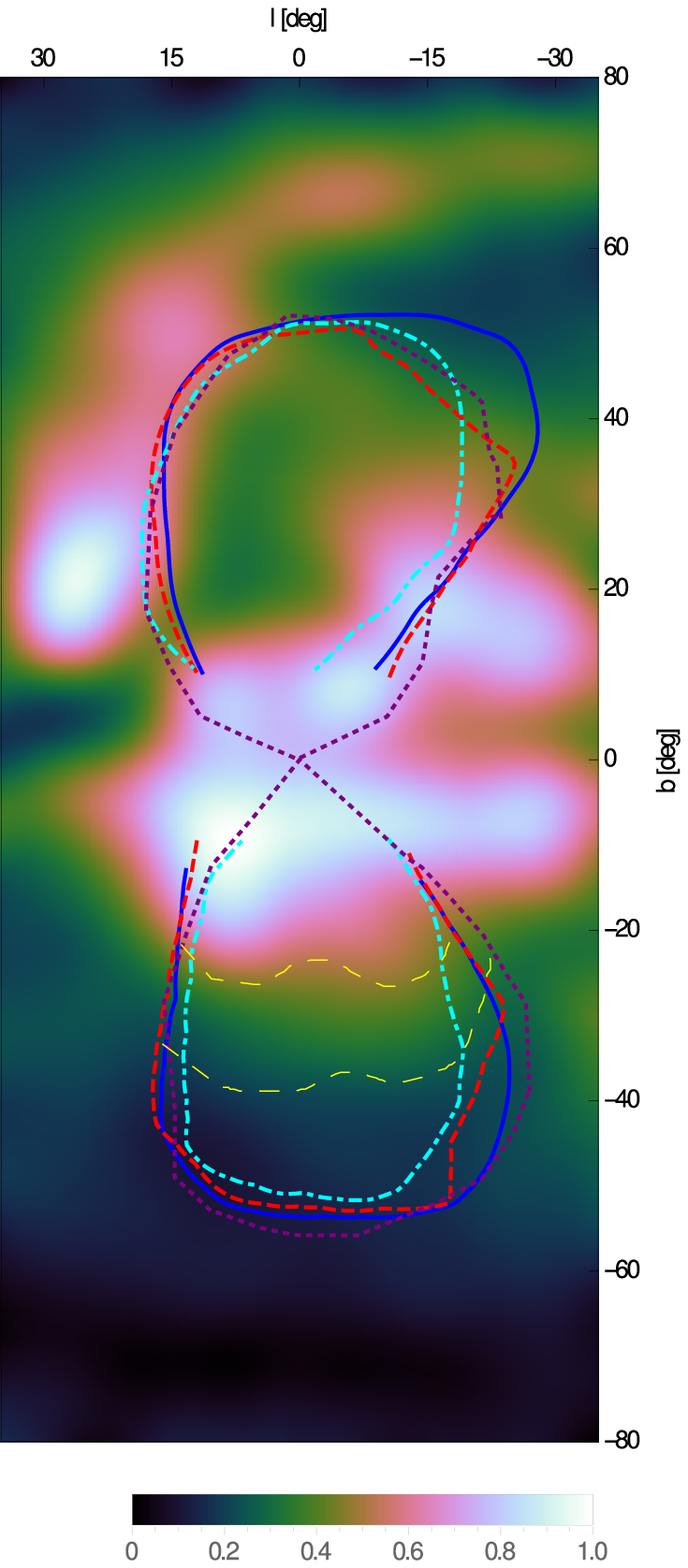}}
 \put (1,91) {\normalsize \textcolor{white}{(d)}}
\end{overpic}
}
}
\caption{
\emph{ROSAT} band R6 ($0.73\keV<E<1.56\keV$) image in Galactic coordinates, with a rectangular (CAR) projection and a cube-helix \citep{Green11_Cubehelix} colormap. Shown are both the raw map (panels $a$ and $b$; color scale: $10^{-6}\mbox{ counts s}^{-1}\mbox{ arcmin}^{-2}$) and the map smoothed with an $8\dgr$ Gaussian filter (panels $c$ and $d$; arbitrary color scale). The four edge contours (see Table \ref{tab:EdgeSummary}) are overlaid in panels $b$ and $d$, as extracted in \citetalias{KeshetGurwich16_Diffusion} based on the Fermi-LAT map with a $6\dgr$ gradient filter (solid blue) and a $4\dgr$ filter (dashed red), or traced by eye in \citetalias{KeshetGurwich16_Diffusion} (dot-dashed cyan) and in \citetalias{SuEtAl10} (dotted purple). Notice that the bipolar, heart shape structure inside the bubbles extends to high latitudes in the south bubble (long-dashed yellow curves in panel $d$), and perhaps also in the west part of the north bubble.
\label{fig:EdgesROSAT6}
}
\end{figure*}

To highlight the association of the bipolar X-ray features with the Fermi bubbles, Figure \ref{fig:EdgesROSAT6} also shows (in panels $b$ and $d$) the FB edges extracted from the \gama-ray data in \citetalias{KeshetGurwich16_Diffusion} and \citetalias{SuEtAl10} (see Table \ref{tab:EdgeSummary}), superimposed on the X-ray map.
The \citetalias{KeshetGurwich16_Diffusion} edges labeled 1 and 2, which we use to extract the stacked, radial profiles in \S\ref{sec:RadialProfiles}, are based on gradient filters of coarse-grained ($6\dgr$) and refined ($4\dgr$) angular scales, applied to the Fermi data.
Also shown are the edges extracted by eye in \citetalias{KeshetGurwich16_Diffusion} (edge 3) and in \citetalias{SuEtAl10} (edge 4).

\MyApJ{\begin{table}[h]}
\MyMNRAS{\begin{table}}
\begin{center}
\caption{\label{tab:EdgeSummary} Different FB edge contour tracing.
}
\begin{tabular}{cll}
\hline
Edge & Tracing method & Reference \\
\hline
1 & Gradient filter on a $6\dgr$ scale & \citetalias{KeshetGurwich16_Diffusion} \\
2 & Gradient filter on a $4\dgr$ scale & \citetalias{KeshetGurwich16_Diffusion}  \\
3 & Traced by eye & \citetalias{KeshetGurwich16_Diffusion}  \\
4 & Traced by eye & \citetalias{SuEtAl10}   \\
\hline
\end{tabular}
\end{center}
\end{table}

For the subsequent analysis, we convert the \emph{ROSAT}/PSPC count rates into physical flux units using the R4--R7 filters in the PIMMS \citep[v4.8d;][]{Mukai93} tool.
For simplicity, the photon flux in each band is converted into the corresponding energy flux in one and the same, $[0.1,2.4]\keV$, wide energy \emph{ROSAT} band, henceforth denoted as $F_X$.
Hence, one may expect the exact same $F_X$ profile to be extracted from the different energy bands, provided that they are dominated by the same signal, with a comparable weighted temperature, and, importantly, that the correct temperature is used in the conversion.
Unabsorbed fluxes are reported, computed using weighted column densities based on the \citet{DickeyLockman90} HI analysis.

We also compute the emission measure, $\mbox{EM}\equiv\int n_e^2 \,dl$, corresponding to $F_X$, with \citep[\eg][]{RybickiBook79}
\begin{eqnarray} \label{eq:EM}
F_X & \simeq & \frac{\alpha c\sigma_T}{\sqrt{6\pi^3}} \int dl \, Z_i^2 n_e^2 \left(\frac{m_e c^2}{k_B T}\right)^{1/2} \int d\epsilon \, \bar{g}_{ff} e^{-\epsilon/k_B T} \nonumber \\
& \simeq & 2.9 \times 10^{-16} \frac{\bar{g}_b}{T_{keV}^{1/2}} \mbox{EM}  \erg \se^{-1}\cm^{-2}\sr^{-1}\coma
\end{eqnarray}
where $\alpha$ is the fine-structure constant, $\sigma_T$ is the Thompson cross section, $c$ is the speed of light, and $m_e$ is the electron mass.
Here, we neglected temperature and metallicity variations along the line of sight, took the cosmic value of the mean squared atomic number, $Z_i^2\simeq 1.2$, and defined $\bar{g}_b$ as the integral (in the first line of Eq.~\ref{eq:EM}) of the weighted Gaunt factor $\bar{g}_{ff}e^{-\epsilon/k_B T}$ over photon energy $\epsilon$. We use the Gaunt factor approximations \citep[\eg][]{DeWittDeWitt73}
\begin{equation} \label{eq:Gaunt}
\bar{g}_{ff}\simeq \begin{cases}
\sqrt{3k_B T/(\pi\epsilon)} & \mbox{for $\epsilon\gg k_B T$} \, ; \\
(\sqrt{3}/\pi)[\ln(4k_B T/\epsilon)-\gamma] & \mbox{for $\epsilon \ll k_B T$} \coma
\end{cases}
\end{equation}
where $\gamma$ is Euler's constant.

\section{Stacked X-ray profiles}
\label{sec:RadialProfiles}

Next, we measure the profile of the X-ray brightness as a function of a varying angular distance $\psi$ from the edge.
The resulting $F_X(\psi)$ profile can then be compared to the model in Figure \ref{fig:FBModelProj}, testing the presence of a shell and providing an estimate of its parameters, in particular the plasma density and temperature.

In order to pick up the weak, diffuse signal, we analyze wide sectors along the FB, and map the pixels onto $\Delta\psi=2\dgr$ wide bins according to their distance $\psi$ from the edge.
The results do not appreciably change for other resolutions, but the statistical fluctuations become prohibitively large for $\Delta\psi\lesssim1\dgr$.

\subsection{South, high latitude profile}
\label{sec:FirstSector}

Consider first the wide, east+west, high latitude sector in the southern bubble, defined by $b<-30\dgr$.
Its X-ray profile measured with respect to the coarse-grained edge 1 is presented in Figure \ref{fig:ProfileBgt30}.
With the above choice of $\Delta\psi$, each angular bin corresponds to a large solid angle, ranging from $\sim63$ square degrees in the innermost bin, to $\sim121$ square degrees in the bin lying just below the edge, to even larger solid angles outside the edge.
The error bars represent the $1\sigma$ statistical confidence levels of each bin, assuming a Poisson distribution.
They do not include the dispersion in the signal among the pixels within the bin, as this is affected by the spatial non-uniformity of the gas distribution, FB asymmetry, gas clumping, and other effects which are beyond the present scope; the smoothness of the resulting signal indicates that our averaging process is meaningful.
(Even the R7 bump around $\psi=3\dgr$ is resolved at smaller $\Delta\psi$.)

\MyApJ{\begin{figure}[t]}
\MyMNRAS{\begin{figure}}
\PlotFigsA{
\epsfxsize=8.5cm \epsfbox{\myfig{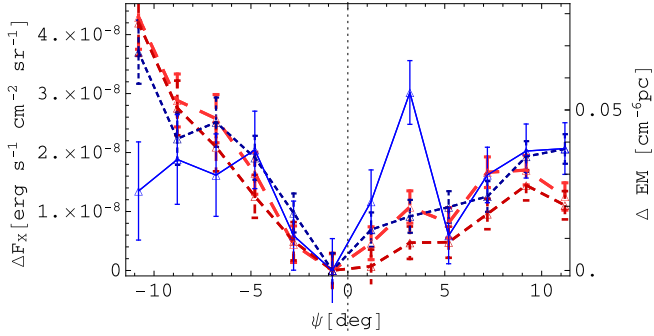}}
}
\caption{
The X-ray flux in the \emph{ROSAT} $[0.1,2.4]\keV$ energy band, as a function of the angular distance $\psi$ from the FB coarse-grained edge 1, for the southern, $b<-30\dgr$, east+west sectors.
Negative $\psi$ values correspond to the inner part of the bubble, \ie closer to the GC.
Plotted is the flux difference, $\Delta F_X$  (left axis), with respect to the bin just inside the edge.
The flux is computed based on the four energy bands R4--R7 (thick, long-dashed, red curve, to thin, solid, blue curve; higher energy bands shown with increasingly thinner lines, shorter dashing, and bluer hue), using the best fit temperature, $k_B T_X=0.4\keV$; see text and Figure \ref{fig:ProfileBgt30T}.
The corresponding emission measure difference ($\Delta\mbox{EM}$; right axis) is computed using Eqs.~(\ref{eq:EM}--\ref{eq:Gaunt}).
\label{fig:ProfileBgt30}
}
\end{figure}

\MyApJ{\begin{figure*}[t]}
\MyMNRAS{\begin{figure*}}
\PlotFigsA{
\centerline{
\begin{overpic}[width=5.8cm]{\myfig{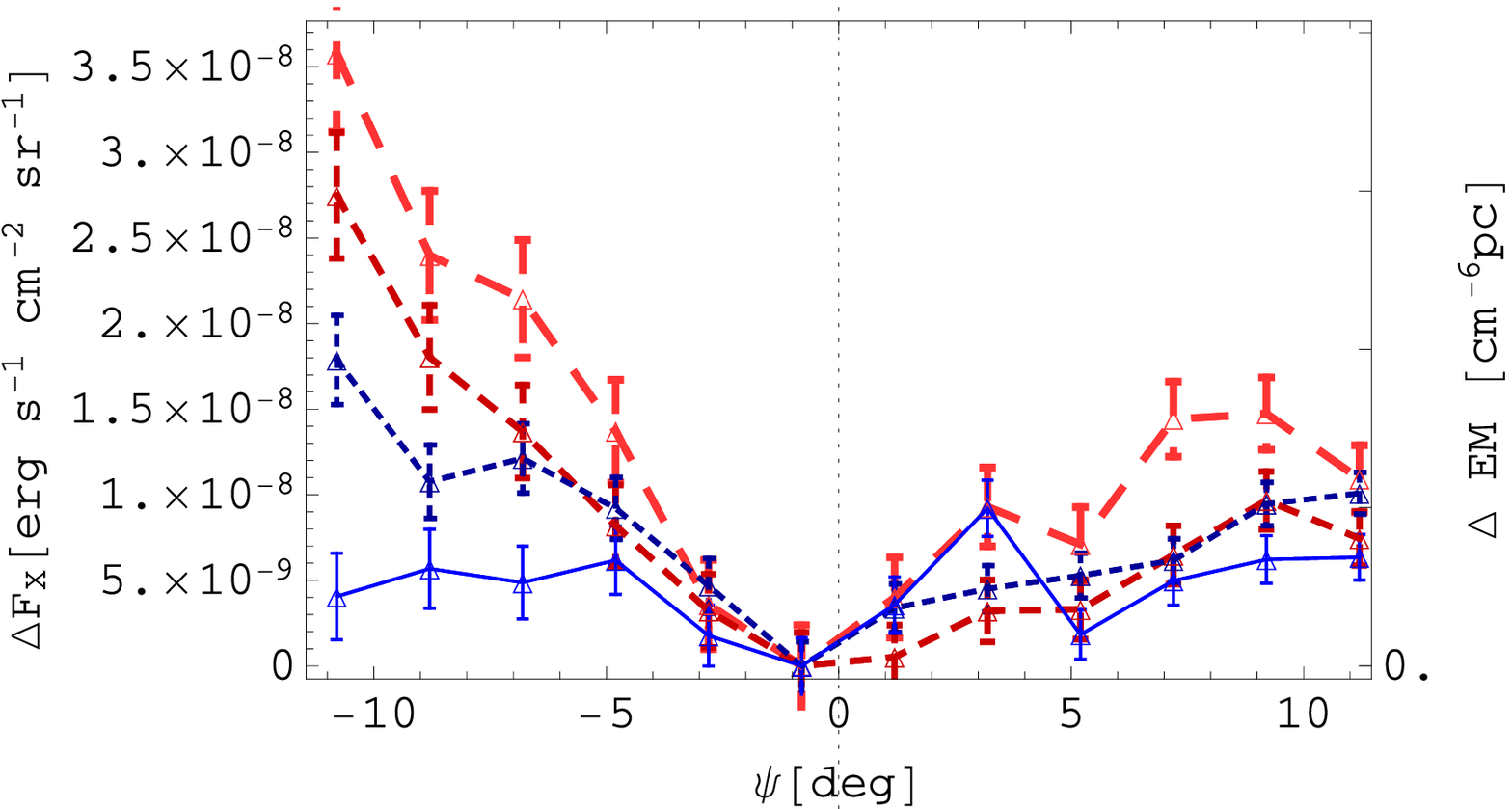}}
 \put (82,46) {\scriptsize \textcolor{black}{(a)}}
\end{overpic}
\begin{overpic}[width=5.8cm]{\myfig{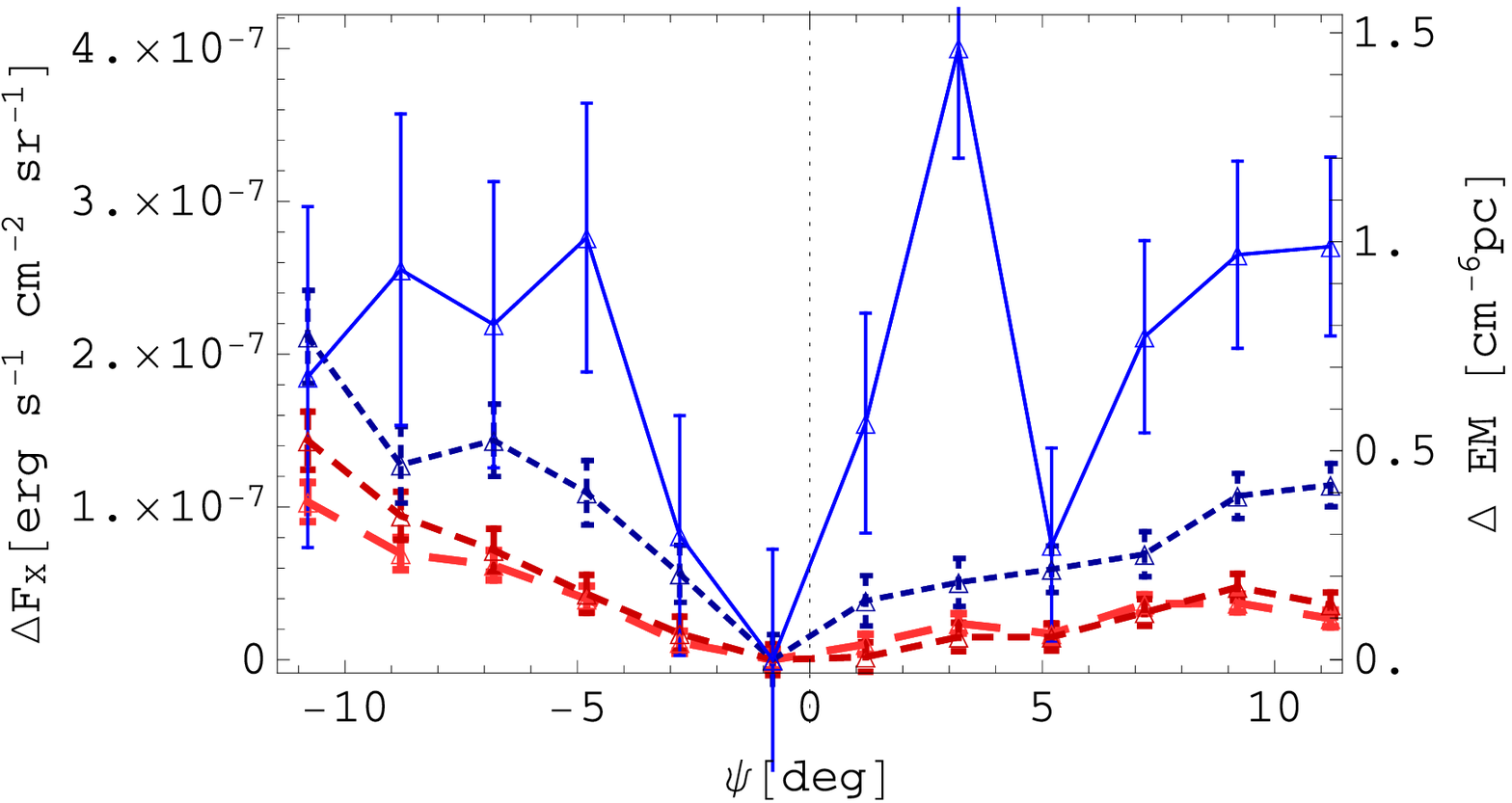}}
 \put (80,47) {\scriptsize \textcolor{black}{(b)}}
\end{overpic}
\begin{overpic}[width=5.8cm]{\myfig{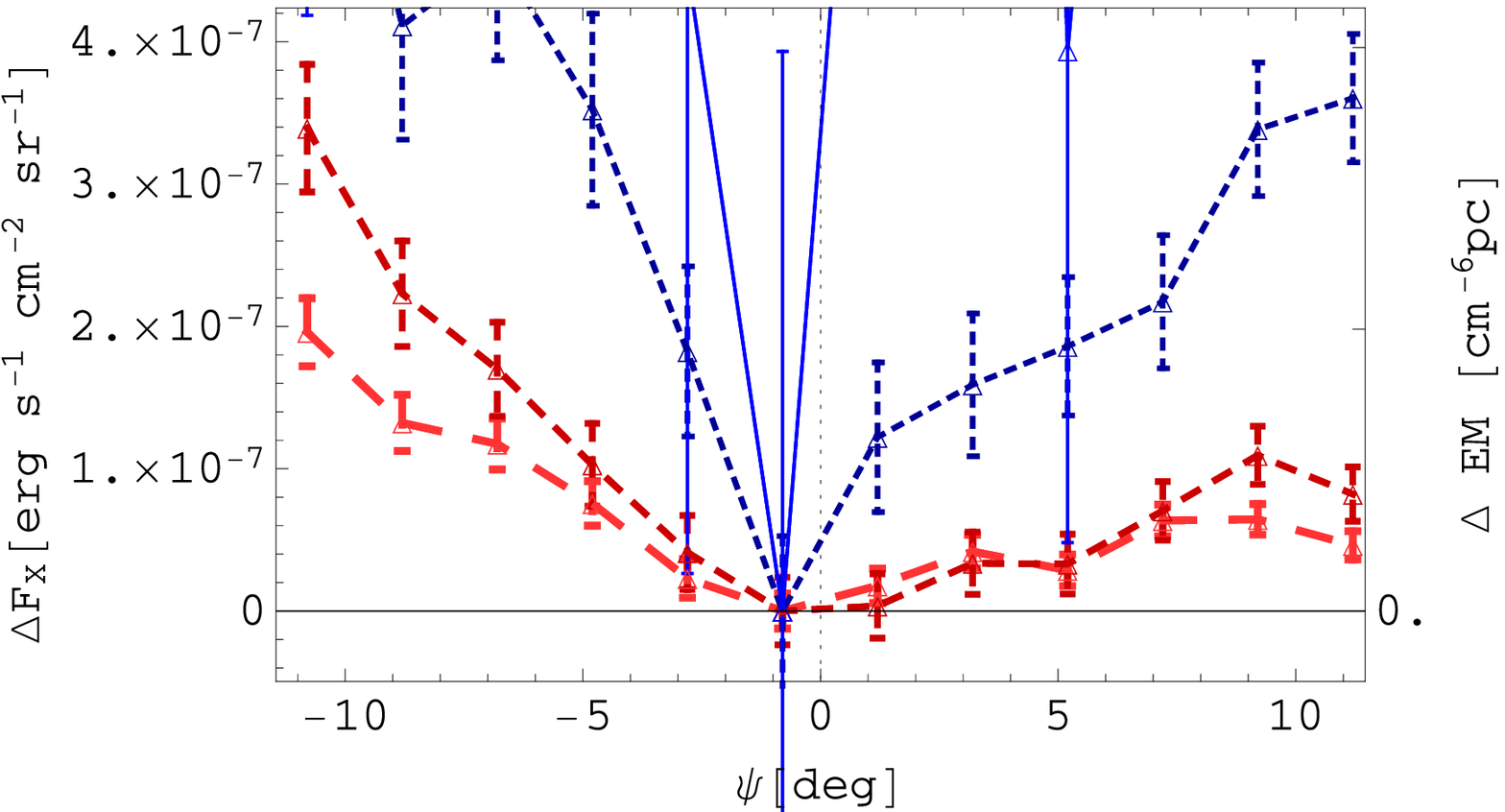}}
 \put (84,49) {\scriptsize \textcolor{black}{(c)}}
\end{overpic}
}
}
\caption{
Temperature dependence of the $b<-30\dgr$ sector shown in Figure \ref{fig:ProfileBgt30} (with the same notations and symbols), shown for $k_B T_X=0.8\keV$ (panel \emph{a}), $0.2\keV$ (panel \emph{b}), and $0.15\keV$ (panel \emph{c}).
The (downstream; $\psi<0$) mismatch between the R4--R6 bands at these temperatures (in both panels \emph{a} and \emph{b}) indicates that $T_X$ lies between $0.2\keV$ and $0.8\keV$;
such $k_BT_X<\keV$ temperature is also consistent with the unclear signal in the high energy band R7; see Figure \ref{fig:ProfileBgt30} and discussion in the text.
Outside ($\psi>0$) the FB,
energy bands R4 and R5 can be matched with $k_B T_X\simeq 0.15\keV$ (panel \emph{c}), but this cannot be confirmed by R6--R7.
\label{fig:ProfileBgt30T}
}
\end{figure*}

The signal in Figure \ref{fig:ProfileBgt30} shows a clear break at the location of the FB edge, with $F_X$ becoming noticeably stronger inward, in resemblance of the expected signature of the supersonic shell in Figure \ref{fig:FBModelProj}.
(Interestingly, in this sector the signal also strengthens outwards; see discussion below.)
Thus, stacking along the edge allows us to measure the weak, extended signal.
The emission measure is $\mbox{EM}\lesssim 0.02\cm^{-6}\pc$ at $(-4)\dgr<\psi<0$.
As expected, this is somewhat lower than the \emph{Suzaku} \citep{KataokaEtAl13} signal and sensitivity in this region.
Indeed, the small field of view in the \emph{Suzaku} observation \citep[$\sim0.9$ square degrees per CCD;][]{KataokaEtAl13} renders its results sensitive to the substantial variations in foreground and signal along the edge, which are averaged out in our method.

In Figure \ref{fig:ProfileBgt30}, similar signatures are seen in each of the three low energy bands, R4--R6, but the signal is less clear in the high energy band, R7, suggesting that the electron temperature is somewhat lower than $\sim1\keV$.
Indeed, the R4--R6 signals agree with each other for the $k_B T_X\simeq 0.4\keV$ conversion temperature used to prepare this figure.
More precisely, this is the temperature we find far ($\psi\lesssim-7\dgr$) inside the edge, where the signal is strong.
Closer to, yet still inside, the edge, the mismatch between bands R4--R6 and the clearer R7 signal suggest a higher temperature; see also Figure \ref{fig:ProfileBgt30T}.
Notice that the temperature is indeed expected to decline with increasing distance inside the edge, by a factor of $\sim 2$ by $\psi=-10\dgr$; see Figure \ref{fig:FBModelProj}.

Figure \ref{fig:ProfileBgt30T} shows the same sector and edge, but with different temperatures assumed in the flux conversion: (from left to right) $0.8$, $0.2$, and $0.15$ keV.
The mismatch here between bands R4--R6 indicates that $T_X$ is indeed lower than $0.8\keV$ (shown in panel \emph{a}), yet higher than $0.2\keV$ (shown in panel \emph{b}).
We conclude that in this sector, far ($\psi\simeq -10\dgr$) inside the edge, $T_X\simeq 0.4\keV$ to within a factor of $\sim2$.

It is difficult to measure the temperature outside the FB edge, where the signal is weaker and the gas is colder than optimal for our energy bands.
Figure \ref{fig:ProfileBgt30T} shows (in panels \emph{b} and \emph{c}) that bands R4 and R5 are well matched for $k_BT_X\simeq 0.1$--$0.2\keV$; this would place these bands on the exponential decline of the signal.
The rising profile of $F_X$ with increasing $\psi>0$ outside the edge in bands R4--R6 suggests some high energy upstream contamination; see discussion in \S\ref{sec:Analysis}.

\subsection{High and intermediate latitude profiles}

In the above method, we measure the stacked X-ray profiles in ten smaller sectors, at both east and west longitudes, both high and intermediate latitudes, and in both hemispheres.
We use the sectors defined in \citetalias{KeshetGurwich16_Diffusion}, as summarized in Table \ref{tab:EdgeSectors}, labeled by lowercase letters \emph{a} through \emph{e}, with or without a hemispheric designation \emph{N} (north) or \emph{S} (south).
Defining the $\psi=0$ contour according to FB edge 1, which in turn is based on the coarse-grained gradient filter, yields the results shown in Figure \ref{fig:RadialProfilesCG}.
Results for the higher resolution gradient filter (more sensitive to sharp transitions), edge 2, are presented in Figure \ref{fig:RadialProfilesHR}.

\MyApJ{\begin{table}[h!]}
\MyMNRAS{\begin{table}}
\begin{center}
\caption{\label{tab:EdgeSectors} Different sectors along each bubble's edge.
}
\begin{tabular}{ccc}
\hline
Sector & Longitude range & Latitude range \\
\hline
a & $-5\dgr<l<5\dgr$ & $|b|>30\dgr$ \\
b & $l>0\dgr$ & $|b|>30\dgr$ \\
c & $l<0\dgr$ & $|b|>30\dgr$ \\
d & $l>0\dgr$ & $15\dgr<|b|<30\dgr$ \\
e & $l<0\dgr$ & $15\dgr<|b|<30\dgr$ \\
\hline
\end{tabular}
\\
\raggedright Sectors are also denoted by the above notation along with the letter \emph{N} (for northern hemisphere) or \emph{S} (southern hemisphere).
\end{center}
\end{table}

\MyApJ{\begin{figure*}[h]}
\MyMNRAS{\begin{figure*}}
\PlotFigsA{
\centerline{
\begin{overpic}[width=5.7cm]{\myfig{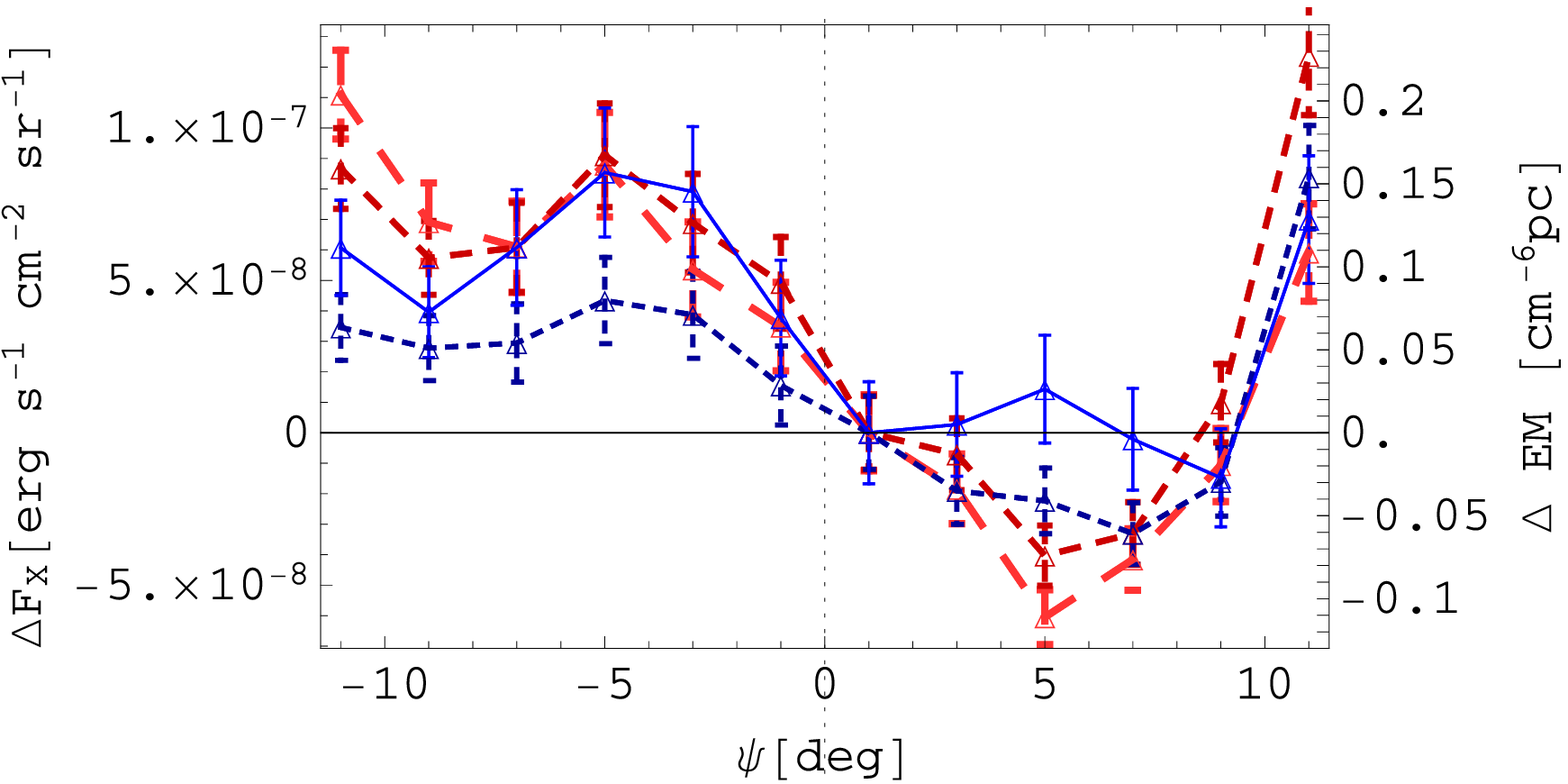}}
 \put (50,44) {\scriptsize \textcolor{black}{(a)}}
\end{overpic}
}
\vspace{-2.0cm}
\centerline{
\begin{overpic}[width=5.7cm]{\myfig{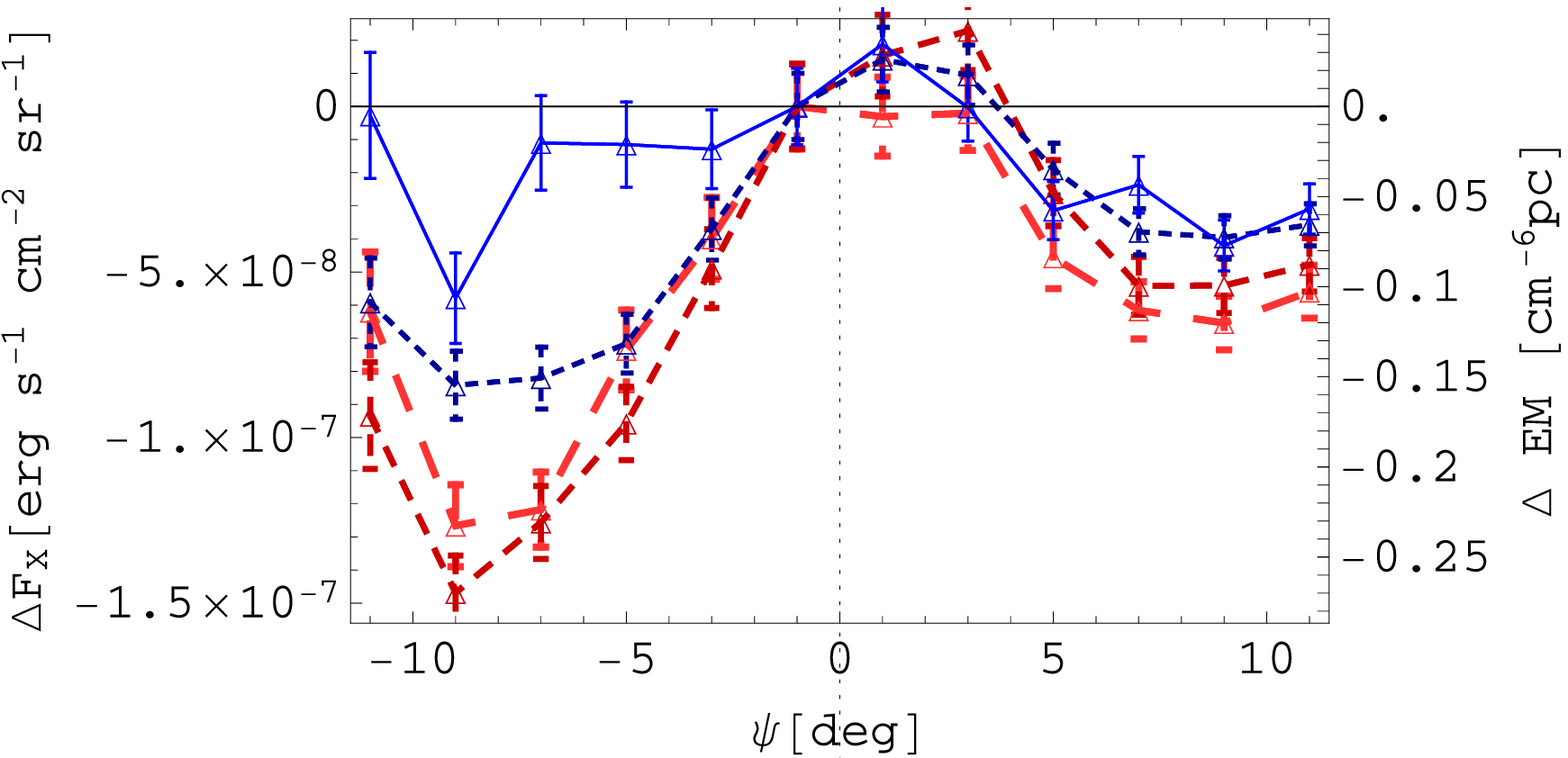}}
 \put (50,36) {\scriptsize \textcolor{black}{(b)}}
\end{overpic}
\hspace{5.7cm}
\begin{overpic}[width=5.7cm]{\myfig{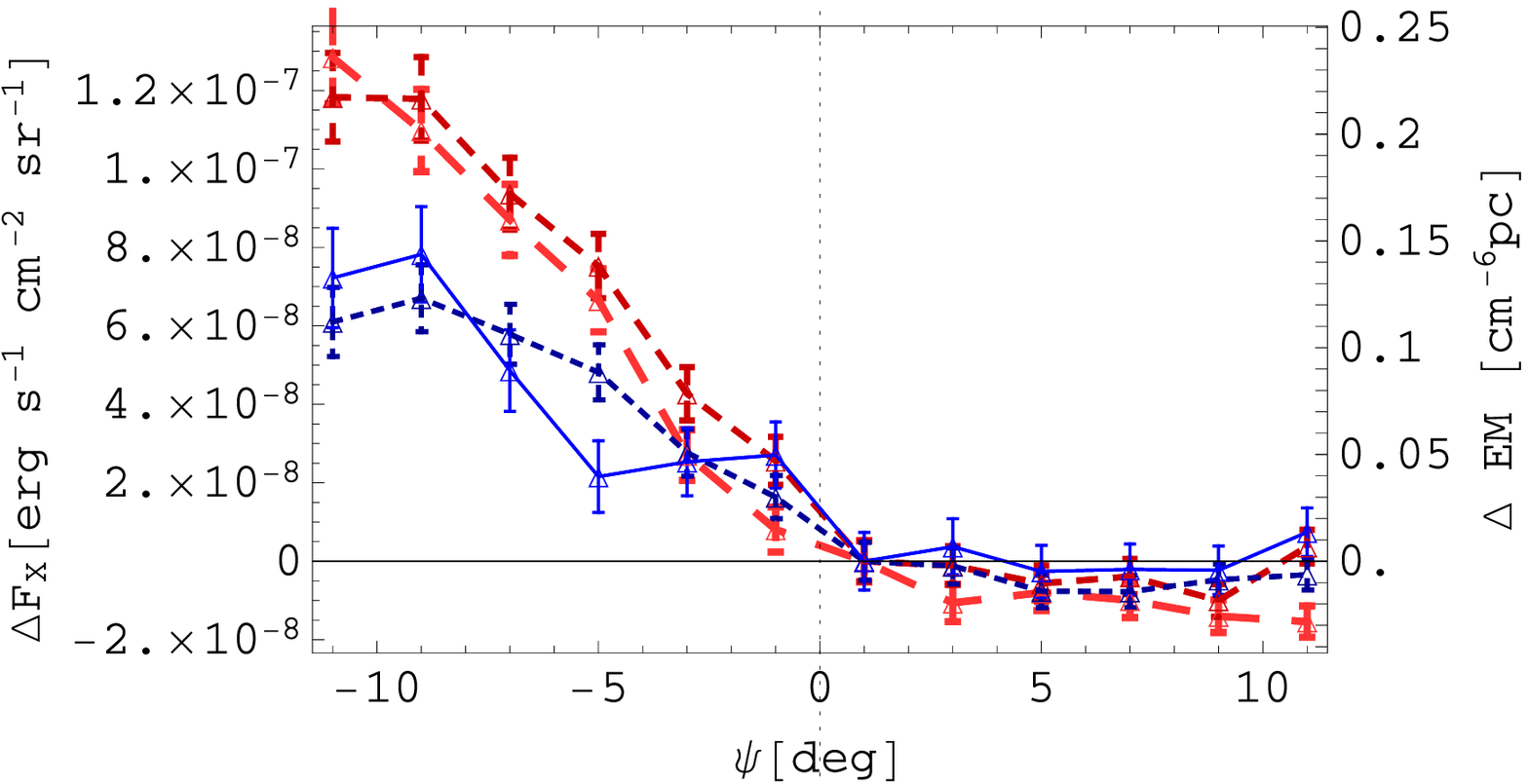}}
 \put (50,44) {\scriptsize \textcolor{black}{(c)}}
\end{overpic}
}
\vspace{0.3cm}
\centerline{
\begin{overpic}[width=5.7cm]{\myfig{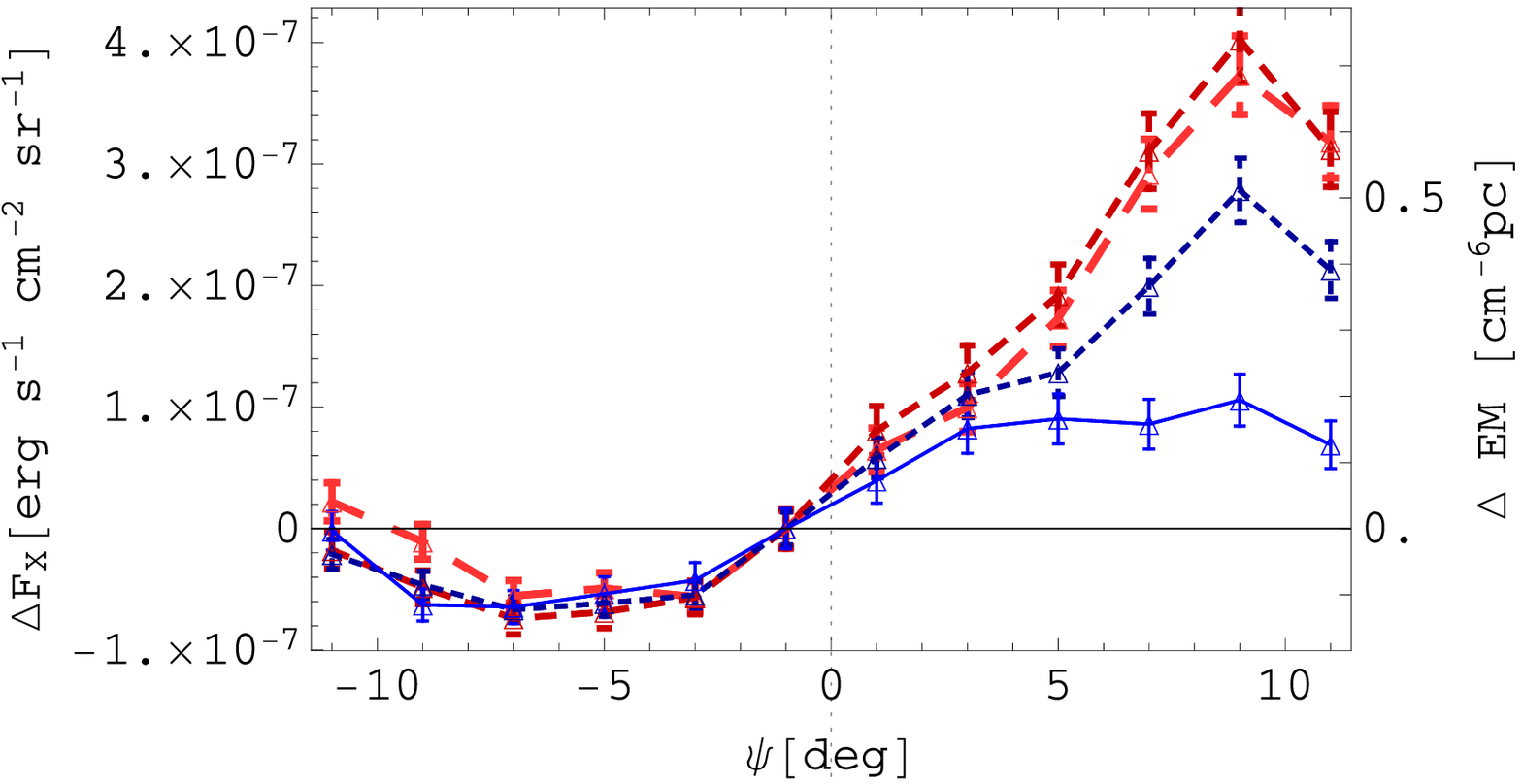}}
 \put (50,46) {\scriptsize \textcolor{black}{(d)}}
\end{overpic}
\begin{overpic}[width=5.7cm]{\myfig{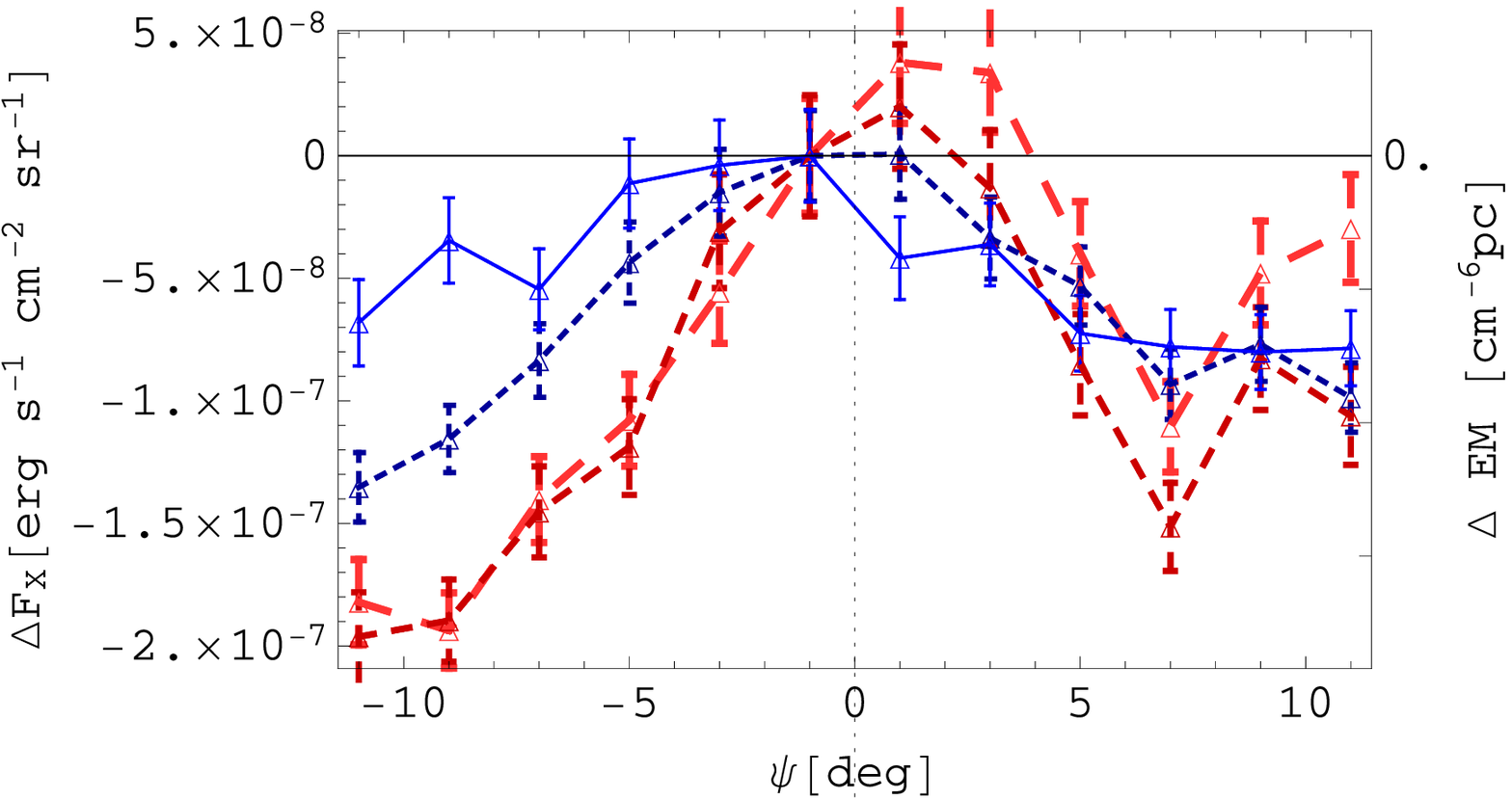}}
 \put (50,48) {\scriptsize \textcolor{black}{(e)}}
\end{overpic}
}
\centerline{
\begin{overpic}[width=5.7cm]{\myfig{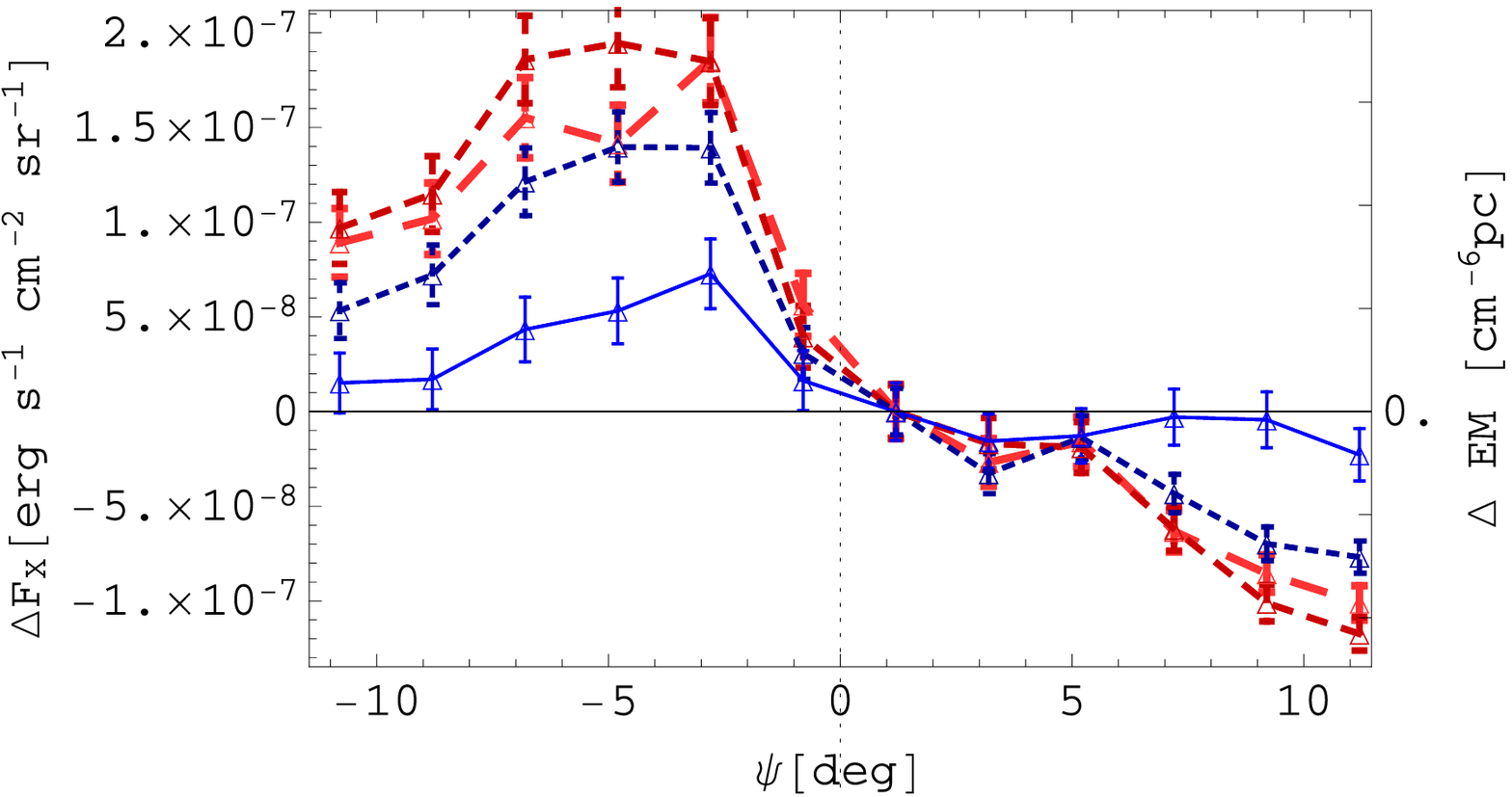}}
 \put (50,46) {\scriptsize \textcolor{black}{(d)}}
\end{overpic}
\begin{overpic}[width=5.7cm]{\myfig{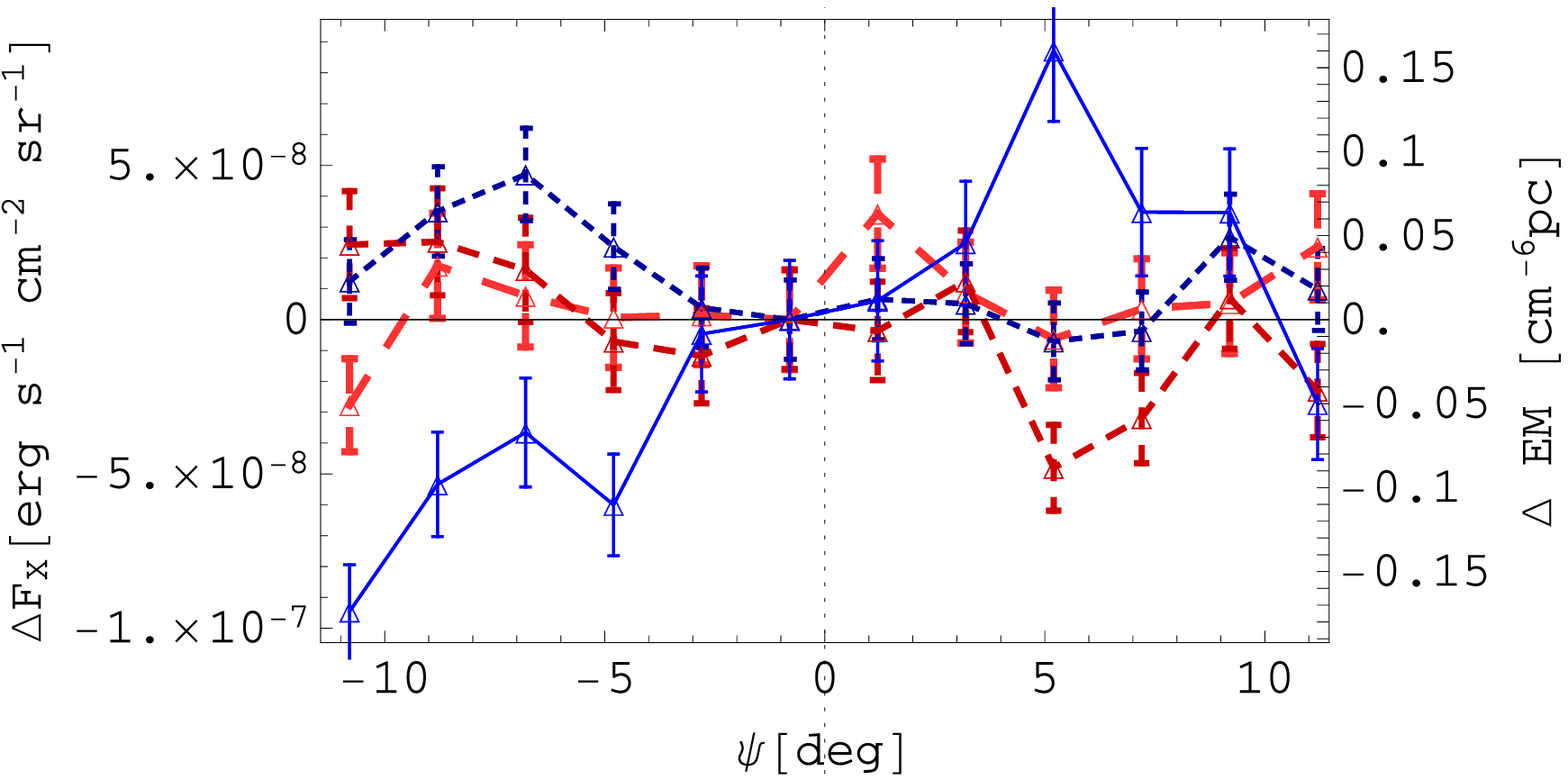}}
 \put (50,44) {\scriptsize \textcolor{black}{(e)}}
\end{overpic}
}
\centerline{
\begin{overpic}[width=5.7cm]{\myfig{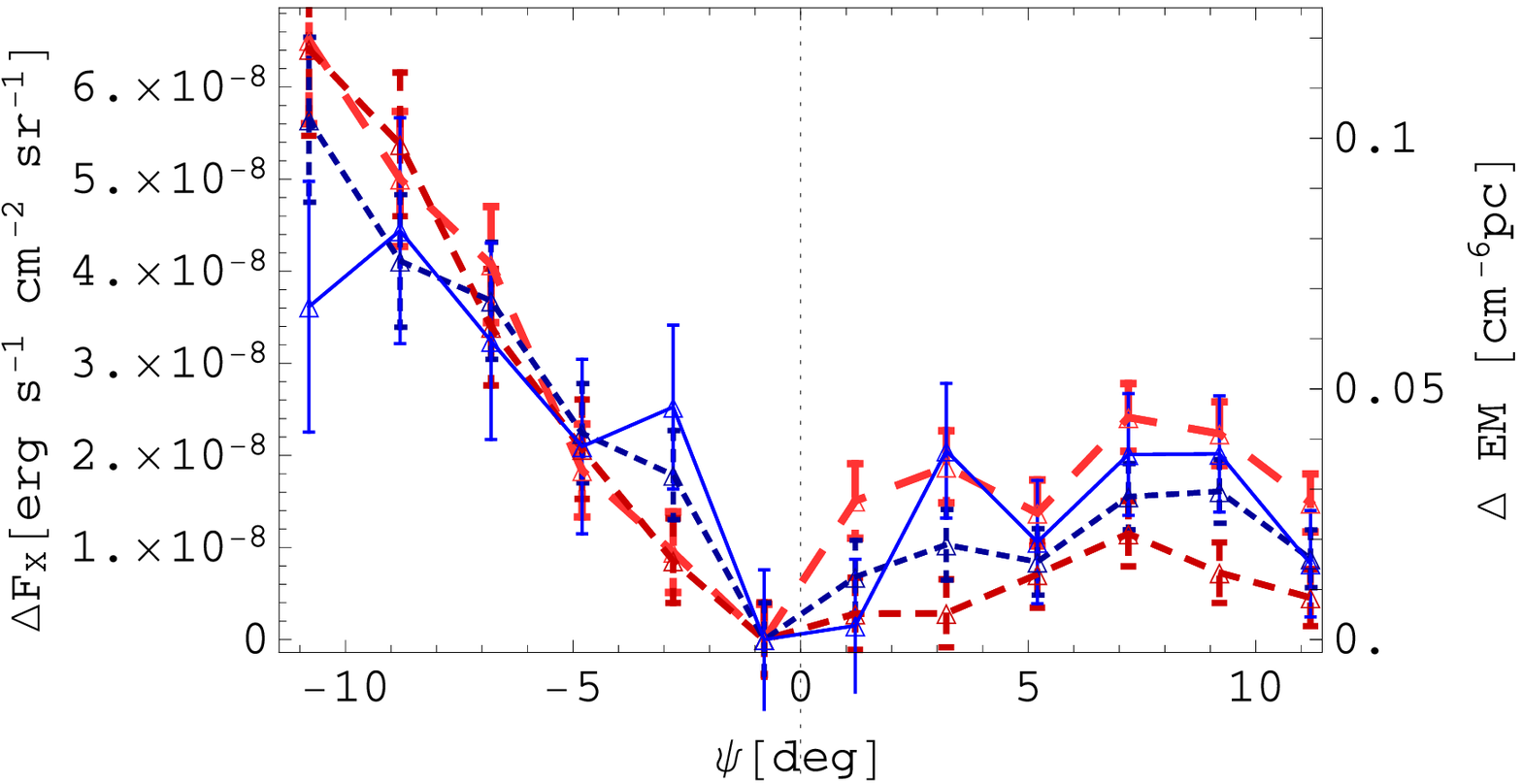}}
 \put (50,44) {\scriptsize \textcolor{black}{(b)}}
\end{overpic}
\hspace{5.7cm}
\begin{overpic}[width=5.7cm]{\myfig{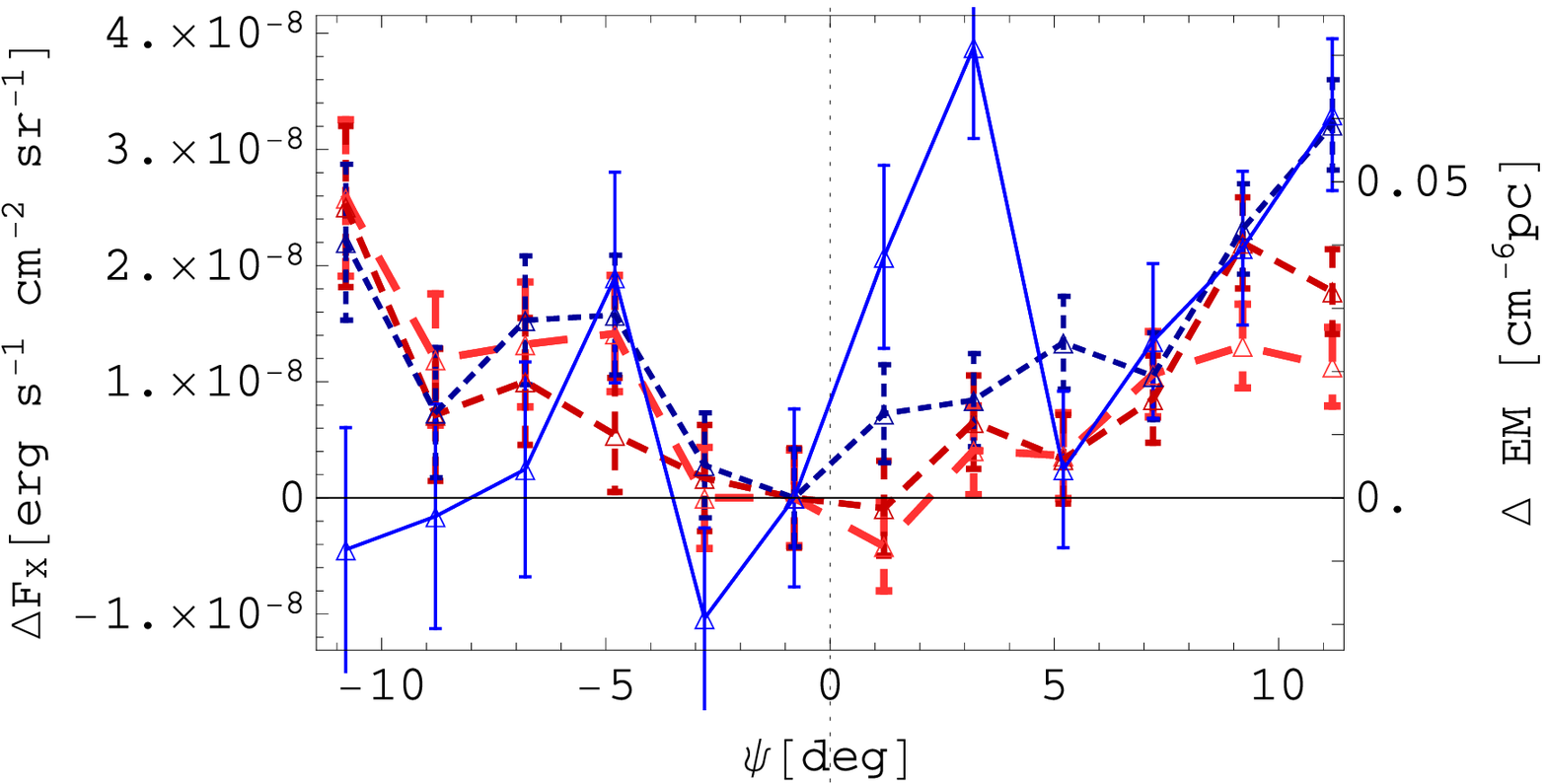}}
 \put (50,44) {\scriptsize \textcolor{black}{(c)}}
\end{overpic}
}
\vspace{-2.0cm}
\centerline{
\begin{overpic}[width=5.7cm]{\myfig{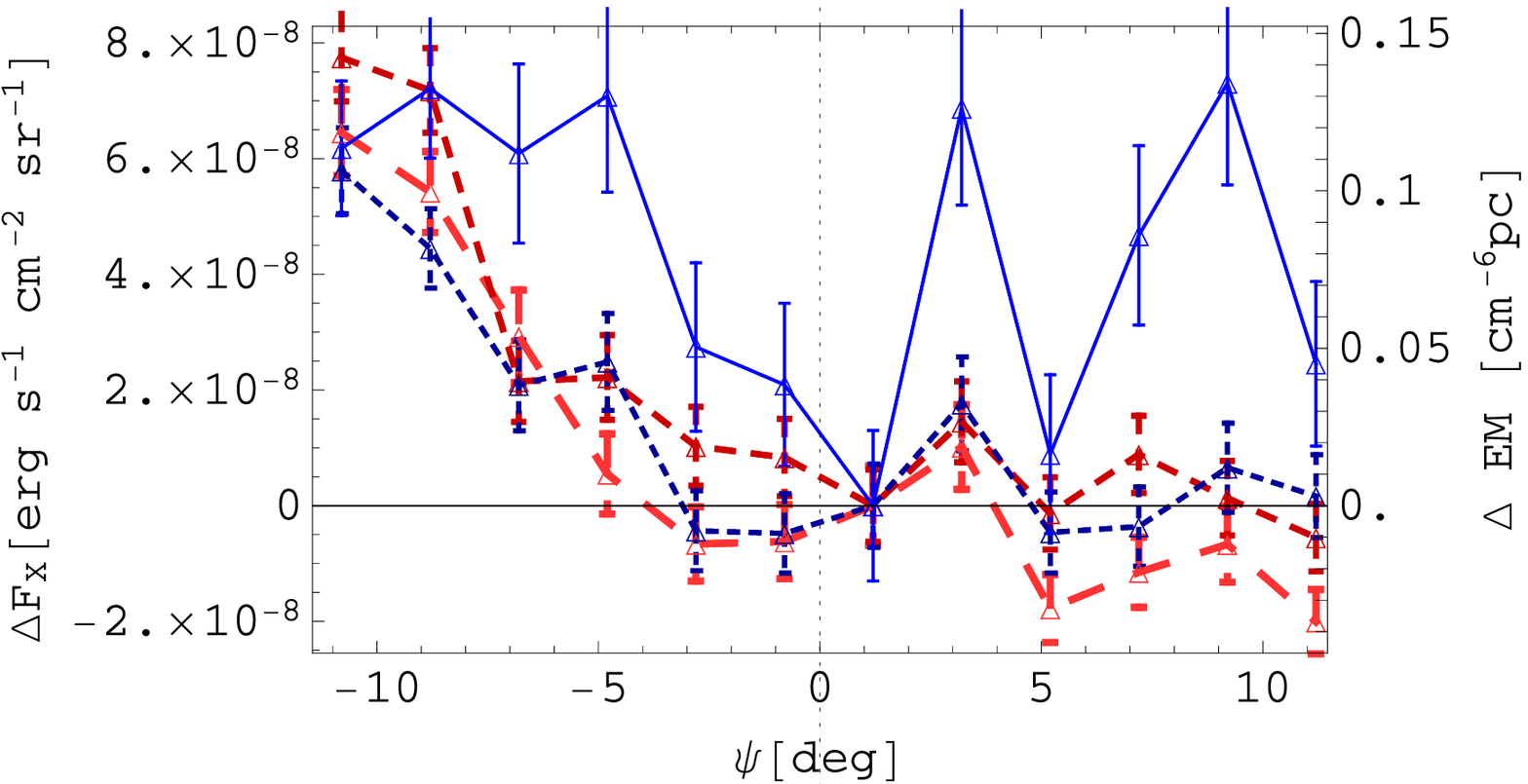}}
 \put (50,44) {\scriptsize \textcolor{black}{(a)}}
\end{overpic}
}
}
\caption{
X-ray flux (and emission measure) difference as a function of angular distance $\psi$ from edge 1
in various sectors (arranged roughly according to their position on the sky; see Table \ref{tab:EdgeSectors}).
We use the conversion temperature $k_B T_X=0.4\keV$, which gives the best fit far below the edge, for all sectors.
Notations and symbols are identical to those used in Figure \ref{fig:ProfileBgt30}.
\label{fig:RadialProfilesCG}
}
\end{figure*}

These figures show the difference of $F_X$ and EM with respect to the FB edge, which can be taken as the first bin either below or above the putative, edge 1 or edge 2 position.
In most cases, we define the edge value according to the bin just below the putative edge, as in Figures \ref{fig:ProfileBgt30}--\ref{fig:ProfileBgt30T}, but some sectors (\emph{aN}, \emph{cN}, \emph{dS}, and for edge 2 also \emph{cS}) yield better results with the first bin above the edge.

For simplicity, we assume a constant temperature within each sector when converting the \emph{ROAST}/PSPC counts to energy flux; the more realistic, $\psi$-dependent temperature is beyond the scope of the present work.
In all sectors that show a signal inside the FBs, the best fit is obtained with $k_BT_X\simeq 0.4\keV$ (up to a factor of $\sim2$) far ($\psi\simeq -10\dgr$) from the edge.
There is evidence for higher temperatures closer to the edge, as discussed in \S\ref{sec:FirstSector} above, but here the statistical errors become large.

\MyMNRAS{\begin{figure*}}
\PlotFigsA{
\centerline{
\begin{overpic}[width=5.7cm]{\myfig{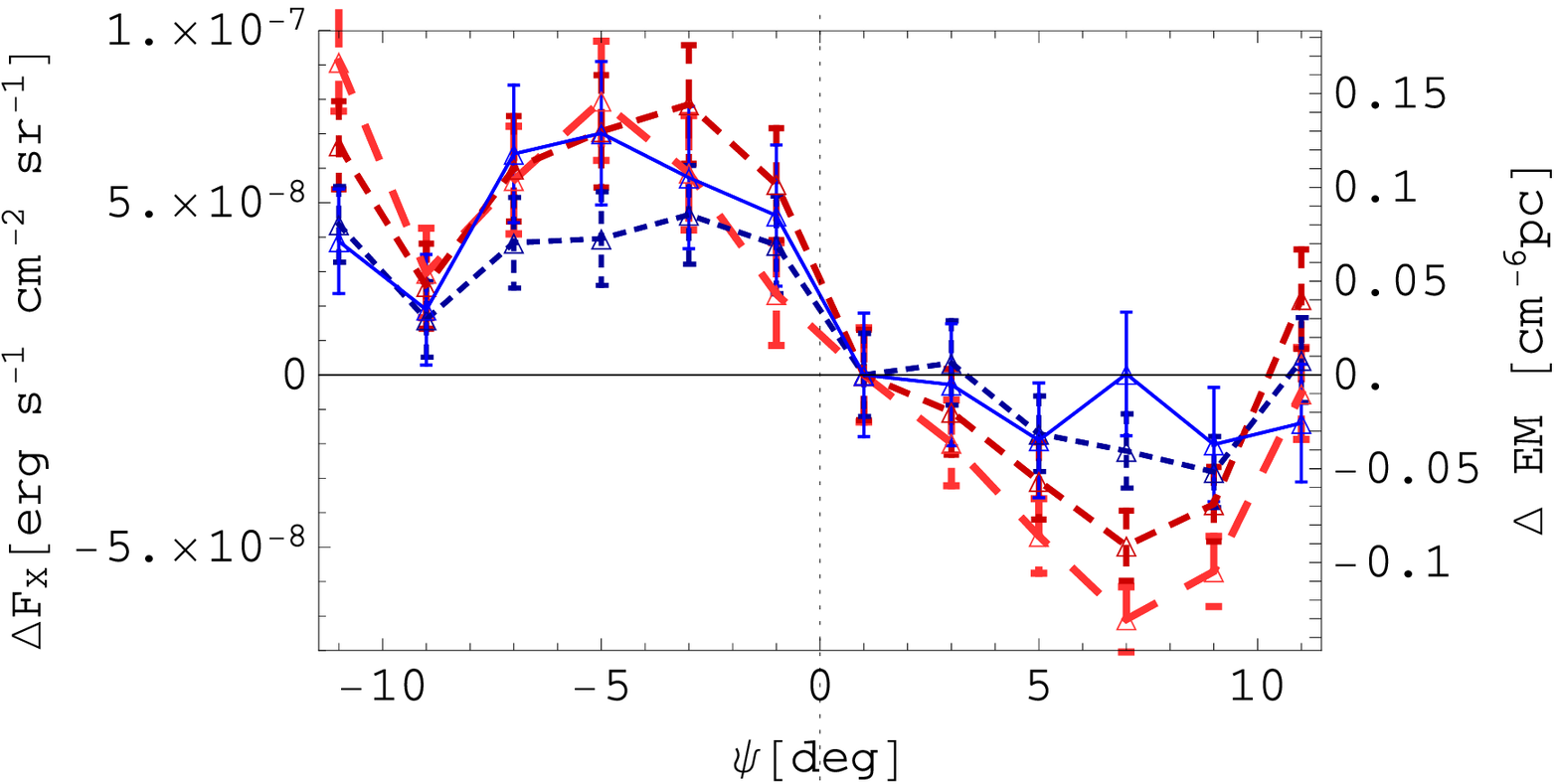}}
 \put (50,44) {\scriptsize \textcolor{black}{(a)}}
\end{overpic}
}
\vspace{-2.0cm}
\centerline{
\begin{overpic}[width=5.7cm]{\myfig{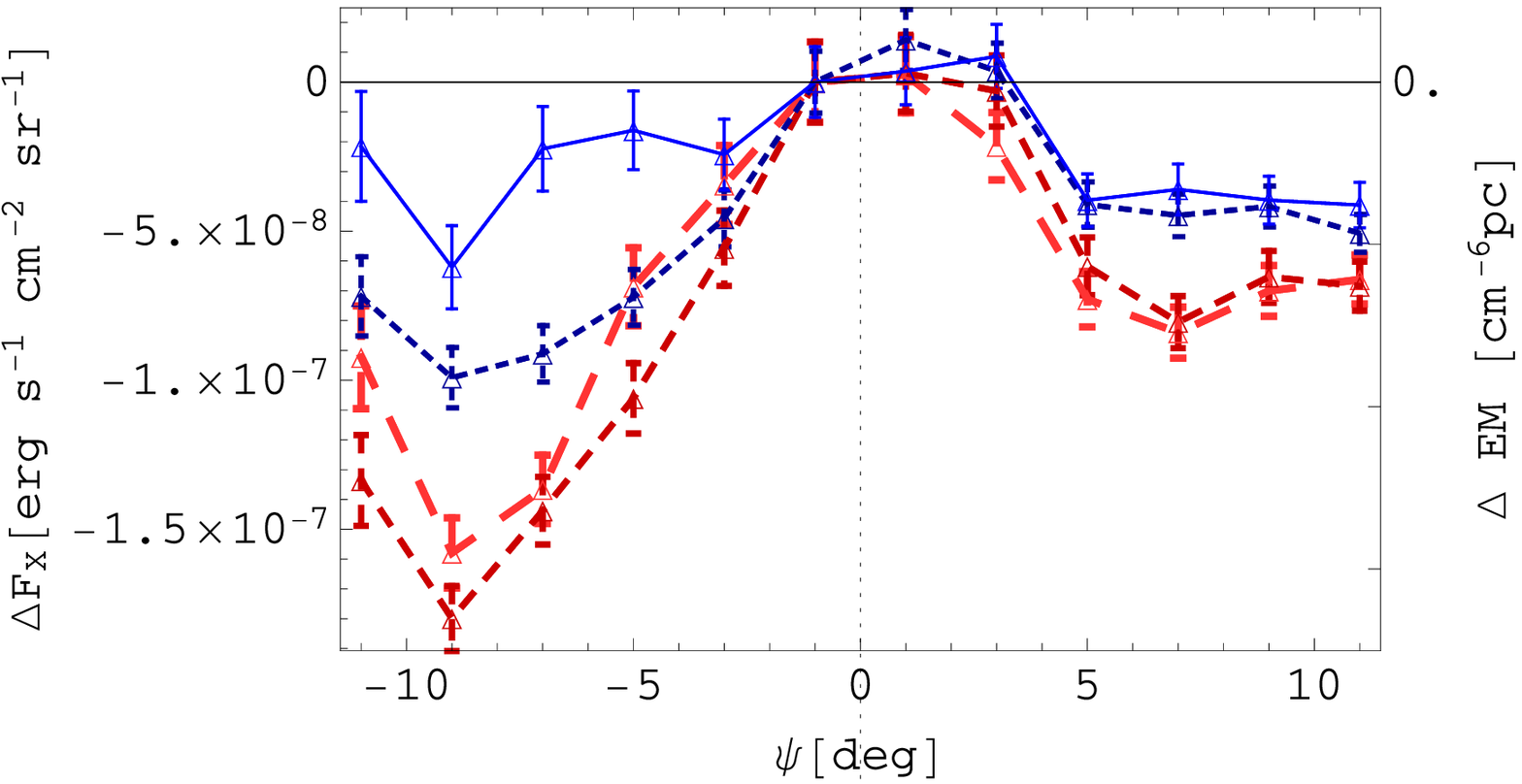}}
 \put (50,47) {\scriptsize \textcolor{black}{(b)}}
\end{overpic}
\hspace{5.7cm}
\begin{overpic}[width=5.7cm]{\myfig{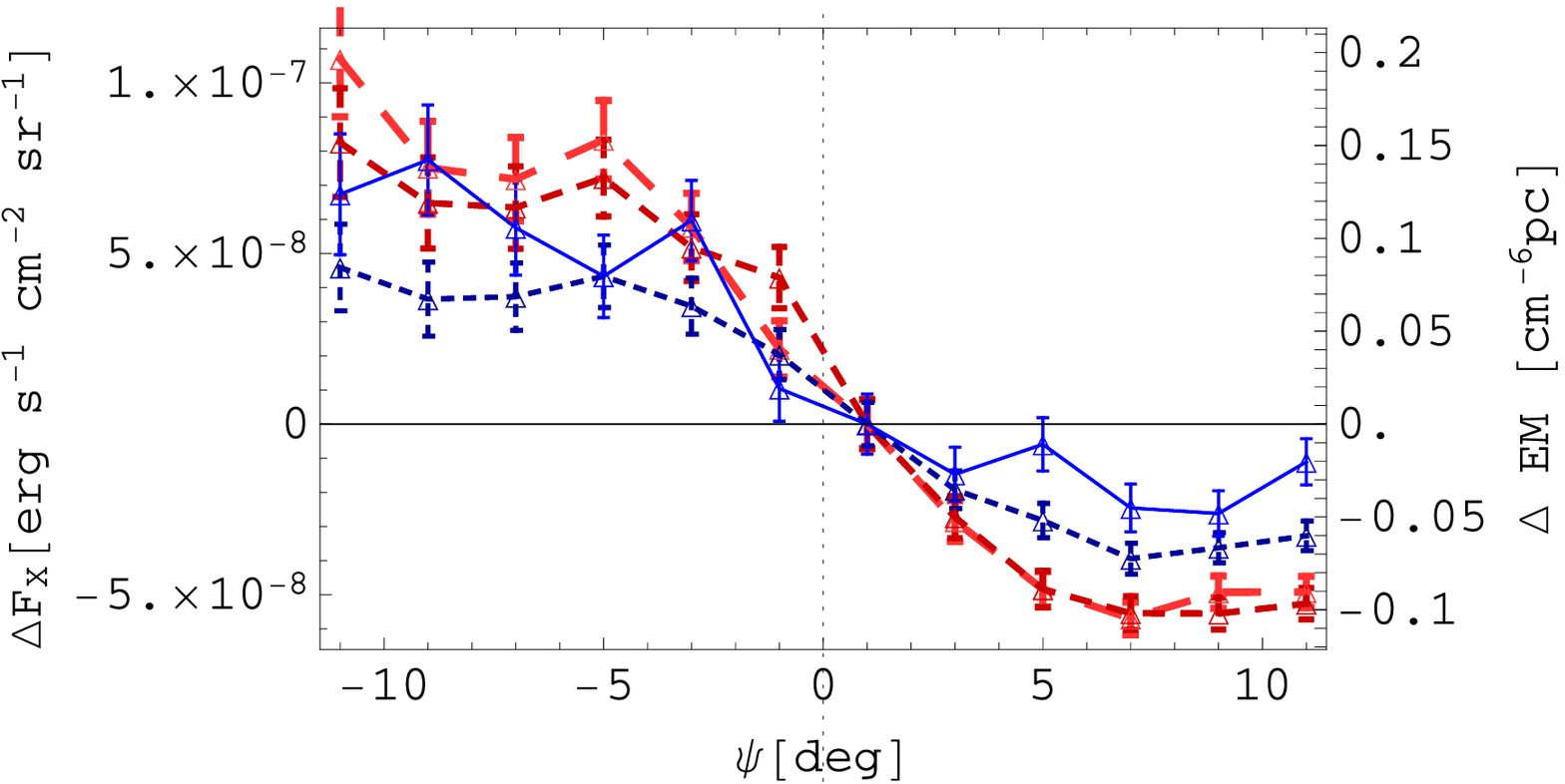}}
 \put (50,44) {\scriptsize \textcolor{black}{(c)}}
\end{overpic}
}
\vspace{0.3cm}
\centerline{
\begin{overpic}[width=5.7cm]{\myfig{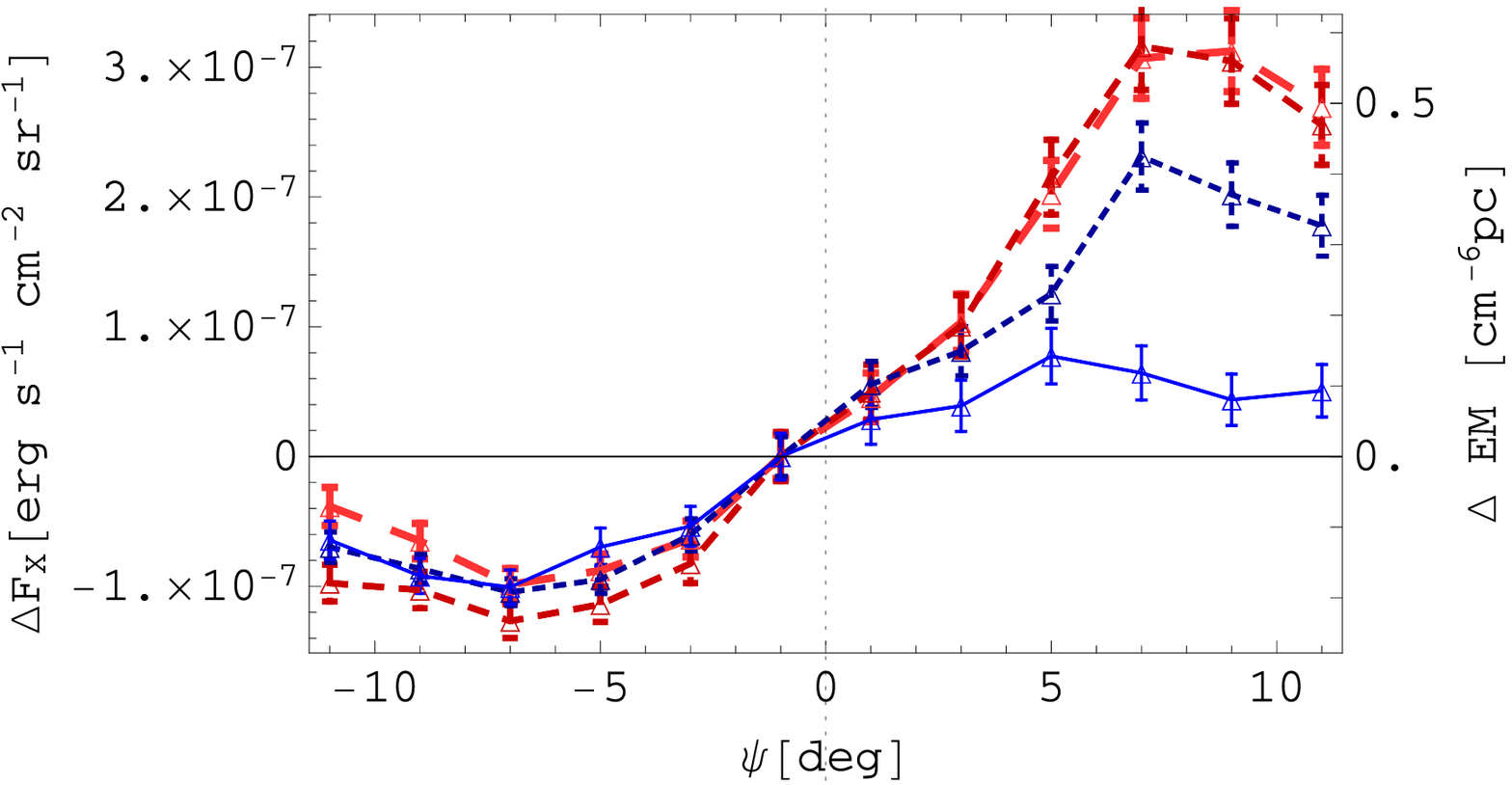}}
 \put (50,47) {\scriptsize \textcolor{black}{(d)}}
\end{overpic}
\begin{overpic}[width=5.7cm]{\myfig{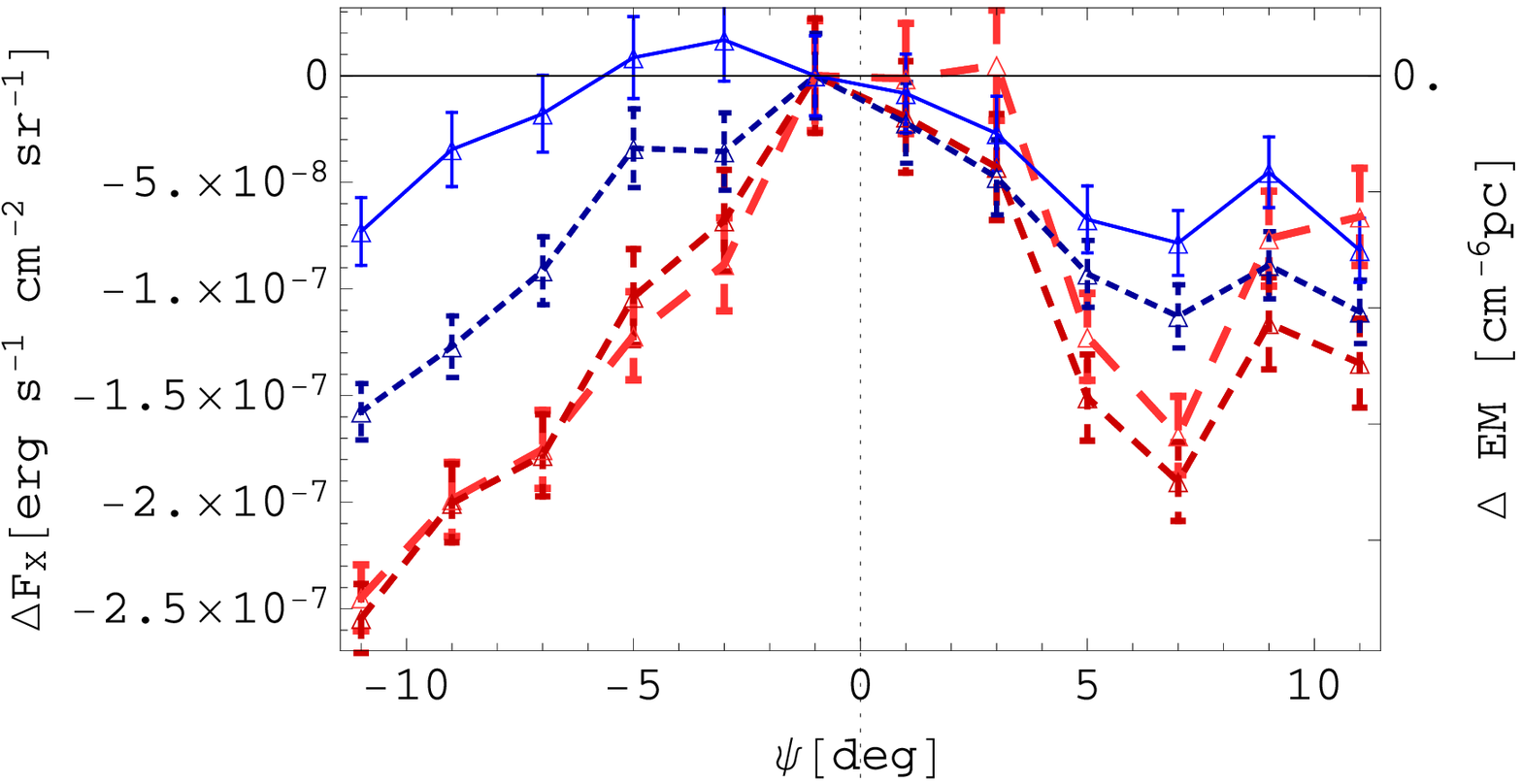}}
 \put (50,37) {\scriptsize \textcolor{black}{(e)}}
\end{overpic}
}
\centerline{
\begin{overpic}[width=5.7cm]{\myfig{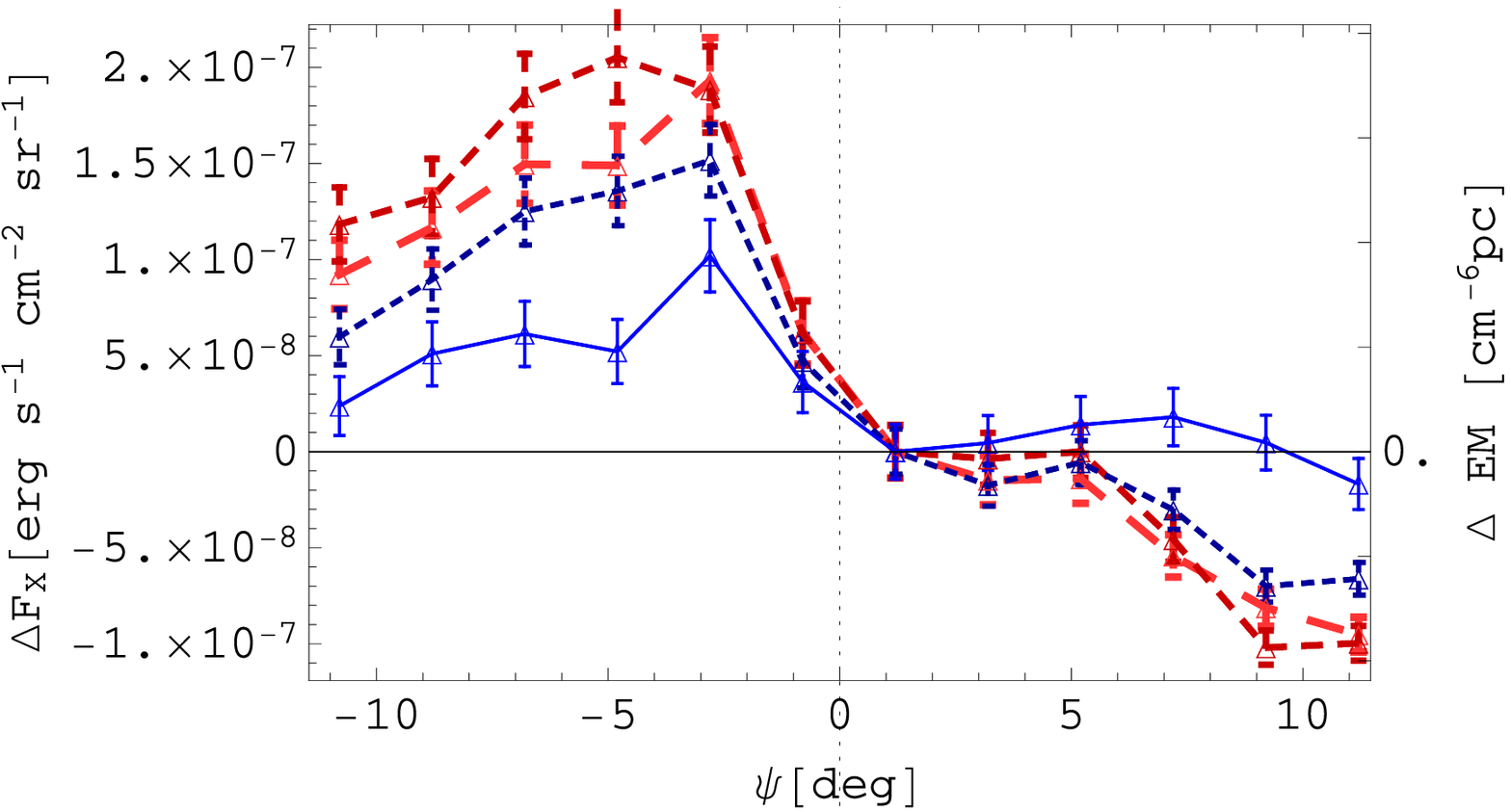}}
 \put (50,47) {\scriptsize \textcolor{black}{(d)}}
\end{overpic}
\begin{overpic}[width=5.7cm]{\myfig{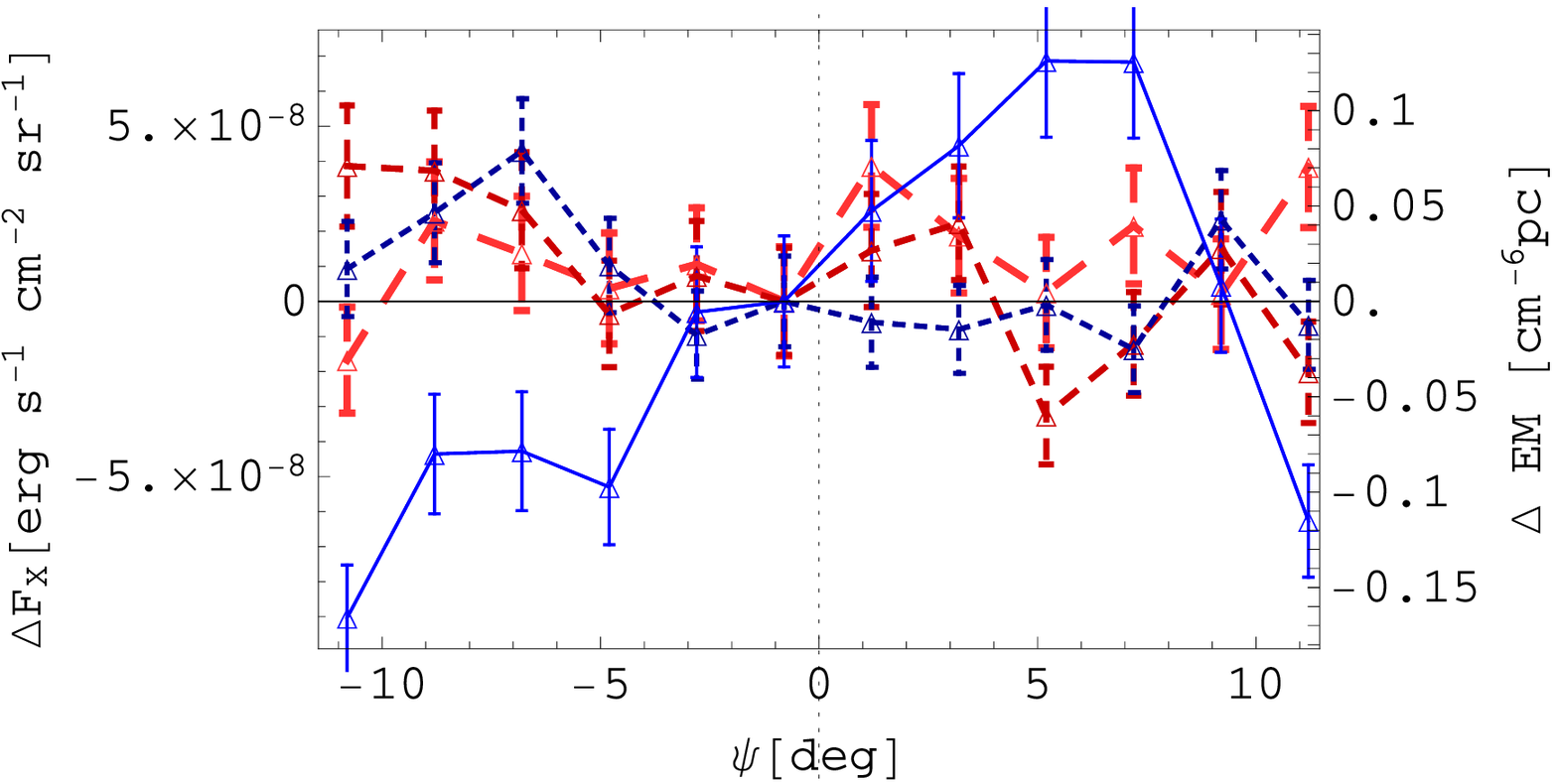}}
 \put (50,44) {\scriptsize \textcolor{black}{(e)}}
\end{overpic}
}
\centerline{
\begin{overpic}[width=5.7cm]{\myfig{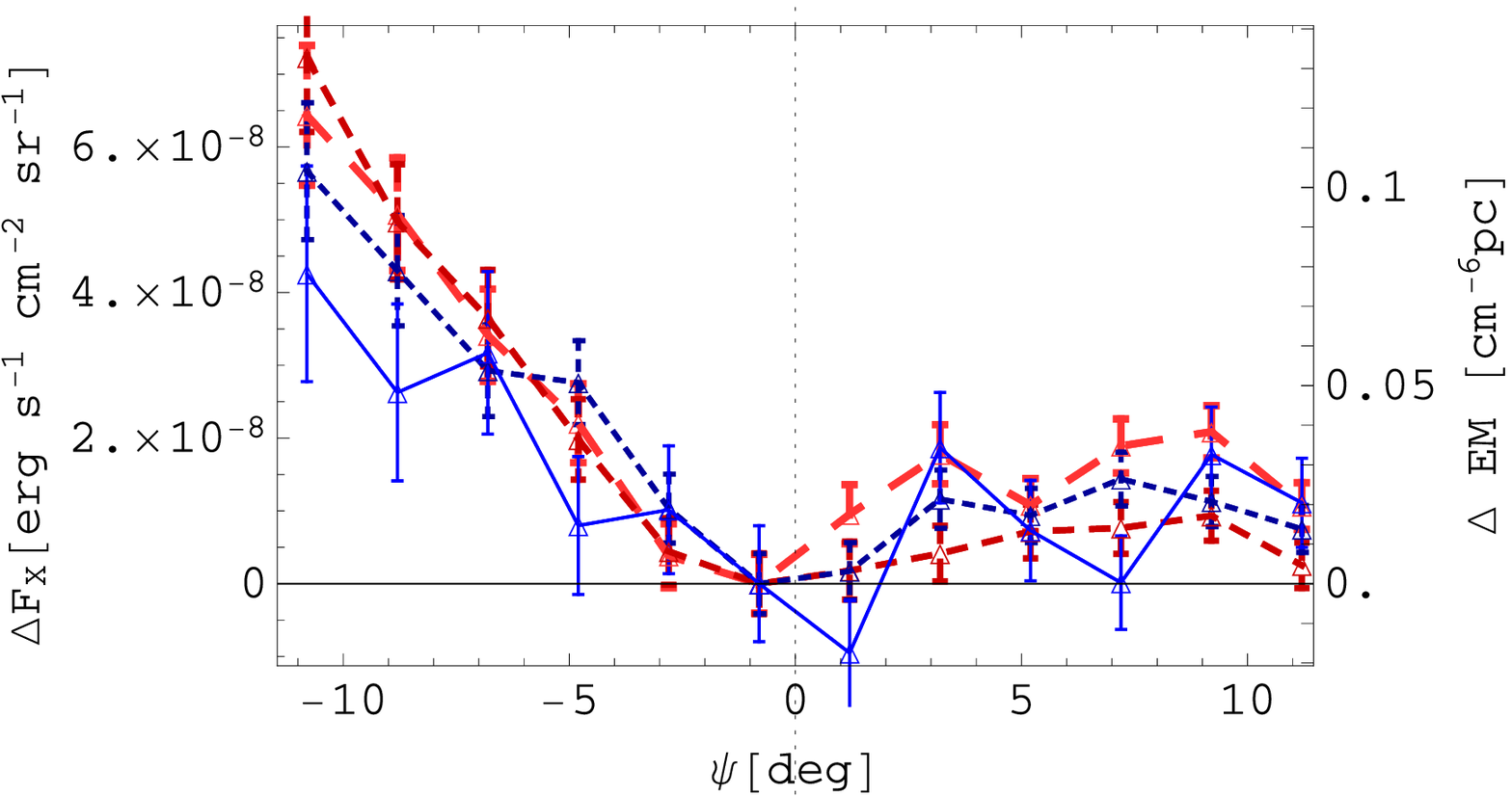}}
 \put (50,47) {\scriptsize \textcolor{black}{(b)}}
\end{overpic}
\hspace{5.7cm}
\begin{overpic}[width=5.7cm]{\myfig{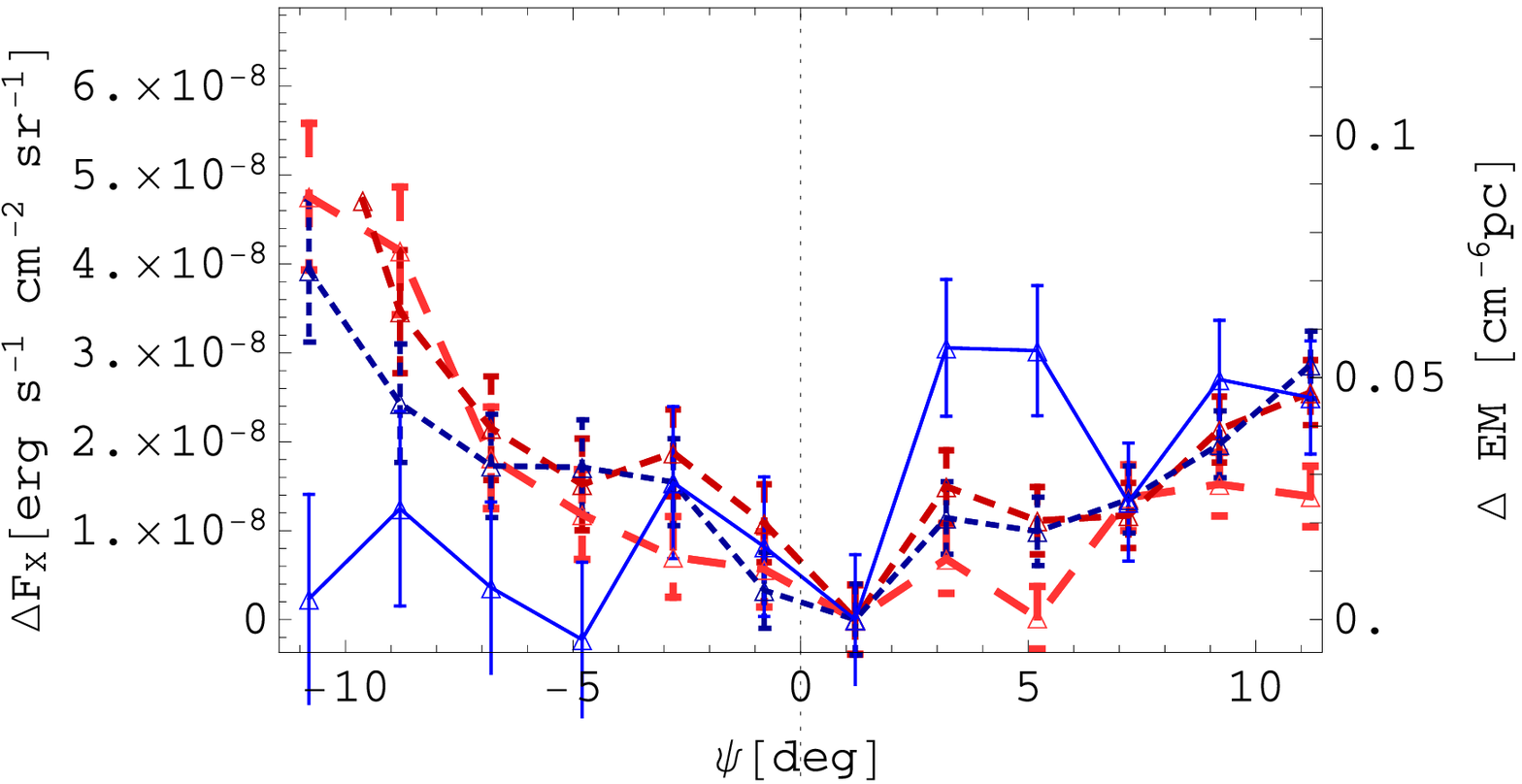}}
 \put (50,47) {\scriptsize \textcolor{black}{(c)}}
\end{overpic}
}
\vspace{-2.0cm}
\centerline{
\begin{overpic}[width=5.7cm]{\myfig{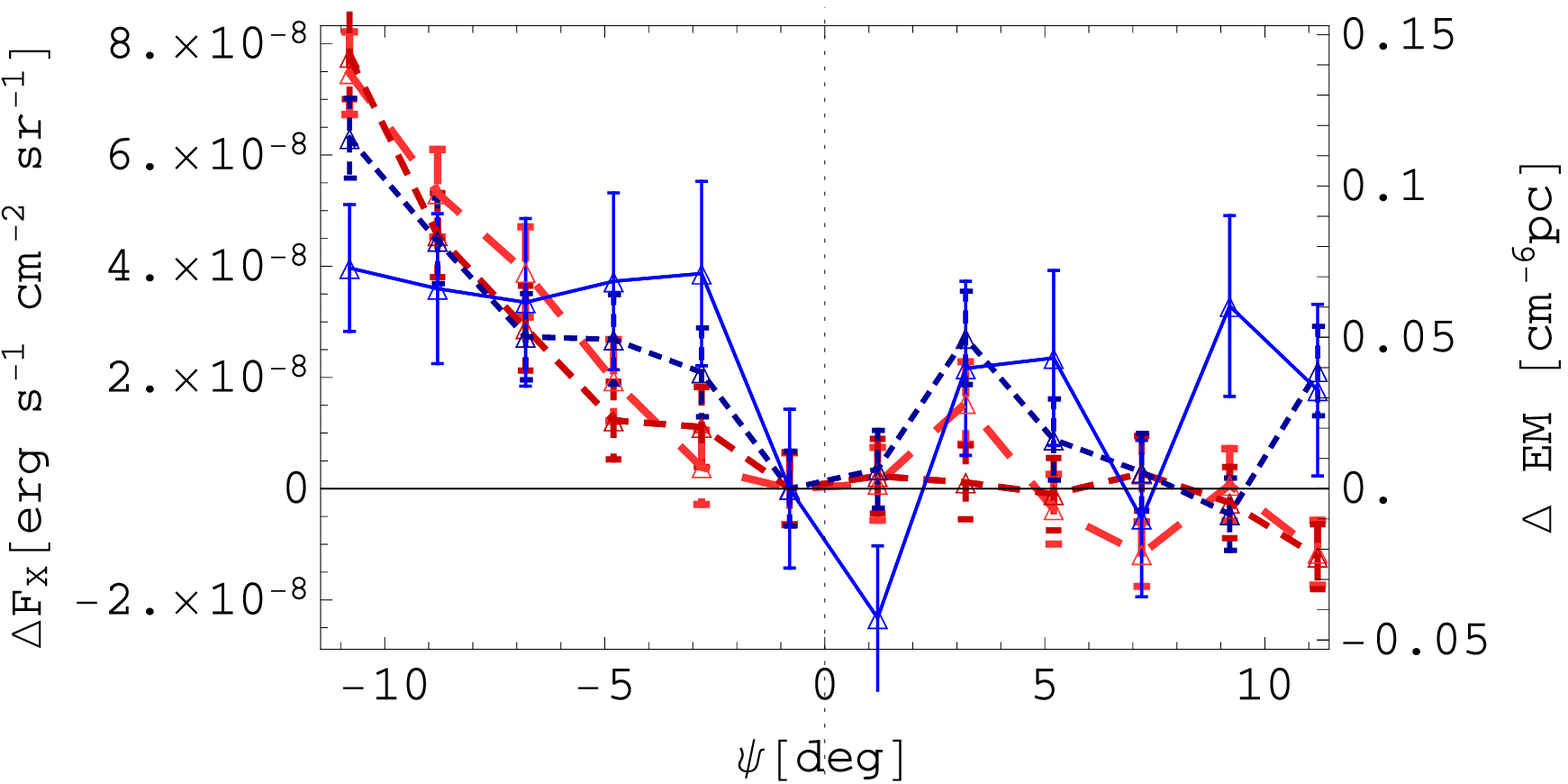}}
 \put (50,44) {\scriptsize \textcolor{black}{(a)}}
\end{overpic}
}
}
\caption{
Same as Figure \ref{fig:RadialProfilesCG}, but for edge 2.
\label{fig:RadialProfilesHR}
}
\end{figure*}

\section{Signal Analysis and modelling}
\label{sec:Analysis}

As Figures \ref{fig:RadialProfilesCG} and \ref{fig:RadialProfilesHR} show, the high latitude ($|b|>30\dgr$) signal in the southern bubble is identified in both the southeast (sector \emph{bS}) and the southwest (sector \emph{cS}), independently. It is also seen if we consider only the bubbles' axis, restricting the analysis to the narrow longitudinal range $-5\dgr<l<5\dgr$ (sector \emph{aS}).
These southern signals are seen when using both edges 1 and 2, with small variations as expected from the differences in the precise edge locations.
We conclude that the signal is robustly confirmed in the southern bubble.

The northern bubble is known to be more prone to confusion, especially near Loop I in the northeast.
Nevertheless, the high-latitude signal can be seen in the north bubble as well, in sectors \emph{aN} and \emph{cN}, albeit not in the northeast sector \emph{bN} which is adjacent to Loop I.
This result, and the similarity between the north and south signatures, especially when using edge 2, support the presence of an underlying X-ray shell associated with both bubbles at high latitudes.

At intermediate ($15\dgr<|b|<30\dgr$) latitudes, only the southeast sector (\emph{dS}) shows clear evidence for the signal, using both edges 1 and 2;
the signal in the adjacent sector \emph{eS} is marginal.
These stacked profiles are considerably more noisy than at high latitudes; no signal is seen in the north. This is to be expected, due to confusion with the abundant X-ray structure near the Galactic plane, and the difficulty of tracing the \gama-ray edges at low-latitude; both effects are more severe in the northern hemisphere.

As mentioned in \S\ref{sec:RadialProfiles}, all high-latitude sectors that show a signal are consistent with $k_BT_X\simeq 0.4\keV$ far ($\psi\simeq -10\dgr$) inside the edge.
We cannot confirm a latitude dependence of $T_X$, but this is not surprising considering the noisy signal at intermediate latitudes and the oversimplified, $\psi$-independent conversion temperature we use in each sector.

As one approaches the FB edge from below, a fairly sharp drop of $F_X$, spanning a few degrees, can be seen in sectors \emph{aS}, \emph{cS}, \emph{dS}, \emph{aN}, and with edge 2, also \emph{cN}.
But these drops are not as sharp and pronounced as in the model Figure \ref{fig:FBModelProj}, and no localized drops are seen in other sectors, in particular \emph{bS}, which is further discussed below.
The excessive smoothness of the measured profiles are likely a result of inaccuracies in tracing the edge position and orientation, along with variations in the actual gas profiles, projection effects, and noise.

The $b<-30\dgr$ signal in Figure \ref{fig:ProfileBgt30} monotonically rises with increasing $(-\psi)$ inward towards the GC.
This resembles the anticipated (\cf Figure \ref{fig:FBModelProj}) signature at intermediate ($15\dgr<b<30\dgr$) latitudes, but is unlike the flattening of the modeled signal at high latitudes, which is seen in Figure \ref{fig:FBModelProj} to be nearly constant for $-10\dgr<\psi<-2\dgr$.
Such unexpected, non-flat behavior is seen in Figure \ref{fig:RadialProfilesCG} to be dominated by sector \emph{bS}; a similar trend is also seen in sector \emph{cN}.
In contrast, the expected flattening of the profile is seen in sectors \emph{aS}, \emph{cS}, and \emph{aN}.
Moreover, this flattening is more pronounced for these sectors when using edge 2 (see Figure \ref{fig:RadialProfilesHR}); here, sector \emph{cN} also shows a clear flattening.
We conclude that the detailed $\psi$-profile is broadly consistent with the model, but cannot be robustly inferred from the present analysis, as it is somewhat sensitive to the method of edge tracing, and may vary across the FBs.
This somewhat diminishes our ability to distinguish between different (presently over-simplified) models for the gas distribution.

Another difference between the measured (high-latitude, southern) profile in Figure \ref{fig:ProfileBgt30} and the model Figure \ref{fig:FBModelProj} pertains to the upstream: the measured signal strengthens away from the edge also with increasing positive $\psi$, outside the FB, instead of being flat or slightly decreasing due to the expected diminishing Galactic emission away from the plane.
This again is seen to be dominated by sector \emph{bS}, although \emph{cS} contributes here as well.
Again, a more consistent, flatter upstream profile is seen in sectors \emph{aS} and \emph{aN}, as well as \emph{cN} and \emph{dS}, and especially when using edge 2.
The unusual profiles both inside and outside the southeast edge \emph{bS} suggest that the upstream gas here differs from other sectors and from our model.
The detection of upstream structure in energy bands R4--R6 and even R7 suggests some high energy upstream contamination in this sector.

In spite of these caveats, we may carry out an approximate, quantitative comparison of the measurements in Figures \ref{fig:ProfileBgt30}, \ref{fig:RadialProfilesCG} and \ref{fig:RadialProfilesHR} with the model Figure \ref{fig:FBModelProj}.
The high latitude profiles reach a flux $F_X(\psi\simeq -10\dgr)\simeq 5\times 10^{-8}\erg\se^{-1}\cm^{-2}$, to within a factor of $\sim2$, whereas the intermediate latitude sector \emph{dS} reaches a flux as high as three times this value.
The corresponding, normalized $n_4^2F_X$ values in the model are comparable to these values, so matching the observations with the model confirms the expected upstream densities.
We conservatively take the discrepancy factor in the flux  to be $D_F\simeq 1$, with an uncertainty factor $\sim 3$, such that the upstream electron number density just outside the top of the bubbles is inferred to be roughly
\begin{equation}\label{eq:nDiscrepancy}
n_e(r=10\kpc) \simeq 4_{-2}^{+4}\times 10^{-4} \left(1D_F\right)^{1/2}\cm^{-3} \fin
\end{equation}
Note that the error bars here and in Eq.~(\ref{eq:TDiscrepancy}) below reflect the variations in the measured and modeled signals, and are not statistical.

In the model, the normalized X-ray temperatures at $\psi=-10\dgr$ are approximately $k_BT_X\simeq 1.8T_{0.15}M_{10}^2\keV$ at high latitudes, and $\sim 0.7\keV$ at intermediate latitudes.
Only the high latitude signal temperature is adequately measured (up to a factor of 2), as $k_BT_X\simeq 0.4\keV$.
The implied $T_X$ discrepancy is a factor of $D_T\simeq1/4$, with an uncertainty factor $\sim 2$, so matching the model crudely yields
\begin{equation} \label{eq:TDiscrepancy}
M_{10}^2 k_B T_u \simeq 0.04_{-0.02}^{+0.04}\left(4D_T\right)\keV \fin
\end{equation}
It is difficult to measure the upstream temperature with bands R4--R7, as illustrated by Figure \ref{fig:ProfileBgt30T}, but temperatures higher than $0.5\keV$ can be excluded.
For a high-latitude temperature $k_BT_u\simeq 0.3\keV$, as found by \emph{Suzaku} \citep{KataokaEtAl13}, the shock Mach number becomes $M\simeq 3.6$, up to an uncertainty factor of $\sim50\%$.
However, lower estimates of the upstream temperature \citep[$k_BT_u\simeq 0.2\mbox{ or }0.15\keV$ according to][and $k_BT_u\simeq0.15\keV$ suggested by Figure \ref{fig:ProfileBgt30T}]{MillerBregman13, MillerBregman16} would imply a stronger, $M\simeq 5$ shock. We conclude that $M\simeq 4$, to within a systematic uncertainty of $\sim2$.

\section{Summary and Discussion}
\label{sec:Discussion}

We analyze the \emph{ROSAT} all sky survey in search of the faint, high-latitude X-ray counterpart of the FB \gama-ray signal.
First, we present a semi-analytic model that reproduces the \gama-ray and low-latitude X-ray signatures of the FBs (see Figures \ref{fig:FBModel} and \ref{fig:FBModelProjGamma}), as well as other constraints, such as the strong shock inferred from microwave and \gama-ray observations, and the absorption line velocities seen towards quasar PDS 456.
This model is then used to compute the signal expected from stacking the \emph{ROSAT} data along the FB edge (Figure \ref{fig:FBModelProj}).
Next, we use the FB edges identified previously \citepalias[by applying gradient filters to the Fermi-LAT map;][see Table \ref{tab:EdgeSummary} and Figure \ref{fig:EdgesROSAT6}]{KeshetGurwich16_Diffusion}, and stack the \emph{ROSAT} data at varying distance from the edge, in various sectors (see Table \ref{tab:EdgeSectors}) along the FBs (Figures \ref{fig:ProfileBgt30}--\ref{fig:RadialProfilesHR}).
The resulting high-latitude signal shows structure clearly associated with the FB edge, in all sectors in the southern hemisphere.
The signal can also be seen in the northern hemisphere, but only in the northwest sectors, far from Loop-I.

Owing to the stacking method, averaging the data over bins of several $10$ square degrees, the statistical errors are rendered manageably small.
Systematic errors due to precise edge localization, projection effects, and competing structure, are more important, but they too are largely washed out in the averaging process.
The similar stacked \emph{ROSAT} signature seen in the different sectors (both north and south, and in the south bubble at both east and west longitudes, and at high and even intermediate latitudes), its approximate agreement with the model predictions, and its robustness against small variations in the edge location and in the analysis parameters (resolution, emission model, absorption model; see below), support a high-significance detection.

The distinguishing characteristic of the signal is the high X-ray brightness found several degrees inside the FBs, declining towards, and dropping as one crosses outside, the FB edge.
This conclusively shows that the FBs are a forward, and not a reverse, shock.
The FBs must therefore arise from a rapid release of energy near the GC, ruling out competing wind or other slow energy release models.
Our results are consistent with the \emph{Suzaku} data \citep{KataokaEtAl13}, showing a similar effect at least in the cleaner, southern hemisphere.

Another important feature of the signal is the $\sim 0.4\keV$ temperature we infer for the emitting electrons far ($\psi\simeq-10\dgr$) inside the edge.
This is evident both from the weak signal in the high \emph{ROSAT} energy band 7 in most sectors, and from fitting the lower energy bands (see Table \ref{tab:ROSATBands}) to the stacked signal.
There is some evidence for a higher temperature closer to the edge and in the highest latitudes (see for example Figures \ref{fig:ProfileBgt30} and \ref{fig:ProfileBgt30T}), but here the data is more noisy.
A radially-increasing temperature inside the FBs, dropping as one crosses outside the edge, is indeed consistent with our forward shock model (see Figures \ref{fig:FBModel} and \ref{fig:FBModelProj}).
The inferred Mach number at the top of the FBs, assuming a thermal equilibrium between shocked electrons and ions, is $M\simeq 4$, with an uncertainty of $\sim2$.

Comparing the stacked results with the projected model in Figure \ref{fig:FBModelProj}, we find that the observed flux and temperature are fractions $D_F\simeq 1$ and $D_T\simeq 1/4$ of their expected values (for our putative model parameters), respectively, with uncertainty factors of $\sim 3$ and $\sim 2$.
This implies similar upstream densities (see Eq.~\ref{eq:nDiscrepancy}) but somewhat lower Mach numbers (Eq.~\ref{eq:TDiscrepancy}) than our fiducial values.
Accordingly calibrating our model, it corresponds to a total energy in (both) the FBs of
\begin{align}
E_{FB} & \simeq 2.4 \times 10^{56} M_{10}^2n_4 T_{0.15} \erg  \\
& \simeq  6 \times 10^{55}  (1D_F)^{1/2} (4D_T) \erg \coma \nonumber
\end{align}
released in a rapid event that took place
\begin{equation}
t \simeq 3.3 M_{10}^{-1}T_{0.15}^{-1/2} \Myr \simeq 6.6 (4D_T)^{-1/2} \Myr
\end{equation}
ago, near \citepalias[within $\sim1\dgr$;][]{KeshetGurwich16_Diffusion} the GC.

These model results should be corrected for the larger extent of the FBs to the west, indicating a higher, $E_{FB}\simeq 10^{56}\erg$ energy.
Importantly, the relatively low $T_X$ we infer is at some tension with the line of sight velocities towards PDS 456, observed to be 3--8 times larger than implied by the calibrated model.
To reproduce these velocities, the model would require a higher gas temperature, and thus would imply younger, more energetic FBs.
A possible resolution of this tension is shock-heating being stronger for ions than it is for electrons.
In particular, $T_i/T_e\simeq 10$ would reconcile the X-ray data with the observed line-of-sight velocities.
This would imply a very strong, Mach $\gtrsim 10$ shock at the top of young, $\lesssim 3\Myr$ FBs, containing a total energy $\sim10^{57}\erg$.
Note that the ion--electron equilibration time would then exceed the age of the bubbles.

The model calibration is based on the clear signals seen several degrees to $10\dgr$ inside the edge; the uncertainty in the radial dependence of the signal dominates our large systematic errors.
Other systematic uncertainties arise from the simplifying assumptions underlying the X-ray analysis (see \S\ref{sec:Data}); in particular approximating the temperature in each sector as fixed.
We tested our results by varying the analysis, for example by replacing the modelled \emph{ROSAT} filters by top-hat filters, and by replacing the absorption column densities by a fixed mean value; the results change within the systematic errors.
Additional systematic uncertainties, not included here, arise from our oversimplified model: we generalized the Primakoff-like spherical kinematics to a bipolar flow, and mostly neglected deviations from axisymmetry (except for a $\sim30\%$ correction to the overall energy budget).

The stacked \emph{ROSAT} signal, like the Fermi-LAT signal, shows no evidence for limb brightening. This indicates that the upstream density declines rapidly with radius; our upstream $n_e\propto r^{-2}$ model (Figures \ref{fig:FBModel}--\ref{fig:FBModelProjGamma}) fits the data much better than an upstream uniform, $n_e\propto r^0$ model (Figure \ref{fig:ST}).
In some sectors, the detailed agreement between model and data is quite good, including the monotonic inward strengthening of the signal at low latitudes vs. the flattening of the signal at high latitudes.
However, this is not observed in all sectors, and is seen (compare Figures \ref{fig:RadialProfilesCG} and \ref{fig:RadialProfilesHR}) to somewhat depend on the precise edge localization.
Accurately inferring the gas distribution underlying the FBs thus requires a more careful tracing of the edge, including deviations from axisymmetry and the consequent projection effects, and cleaning some of the noise, in particular a possible high-energy contaminant upstream of the southeast sector.

Our results constrain some fundamental aspects of the FB phenomenon.
First, the strong shock we deduce is consistent with the spectrum inferred from the microwave haze and from the absence of strong variations in the \gama-ray spectrum along the edge \citepalias{KeshetGurwich16_Diffusion}.
This supports the interpretation of the haze and the \gama-rays as arising from CREs, Fermi-accelerated by the shock.
The very low density we infer rules out hadronic models for the \gama-ray signal, providing strong support for the competing, leptonic models (for a discussion, see \PaperTwo).
While the X-ray signal removes some of the degeneracies in the model, it does not by itself unequivocally prove that the FBs lie at a Galactic distance; the emission measure is quite low \citep[comparable and somewhat lower than reported by][]{KataokaEtAl13}, due to the high latitude.

\MyApJ{\acknowledgements}
\MyMNRAS{\section*{Acknowledgements}}
We thank Y. Lyubarski, R. Crocker, and Y. Naor for helpful discussions.
This research (grant No. 504/14) was supported by the ISF within the ISF-UGC joint research program, and by the GIF grant I-1362-303.7/2016, and received funding from the IAEC-UPBC joint research foundation grant 257.
We acknowledge the use of NASA's SkyView facility (http://skyview.gsfc.nasa.gov) located at NASA Goddard Space Flight Center.

\bibliography{FermiBubbles}



\end{document}